\documentclass[trackchanges,twocolumn]{aastex631}

\RequirePackage{amsthm,amsmath,amsfonts,amssymb}

\usepackage{booktabs} 
\usepackage{multirow}
\usepackage{subfigure}
\usepackage{anyfontsize}
\usepackage{newtxtext,newtxmath}
\usepackage[normalem]{ulem}
\usepackage{enumitem}
\usepackage{algorithm}
\usepackage{algpseudocode}
\usepackage[noabbrev,capitalise]{cleveref}

\usepackage{todonotes}
\setuptodonotes{inline}
\setuptodonotes{color=yellow!30}

\makeatletter
\usepackage{etoolbox}
\patchcmd\H@refstepcounter{\protected@edef}{\protected@xdef}{}{}
\makeatother

\newcommand{\KL}[2]{D_{\rm KL}(#1 \, \Vert \, #2)}

\shorttitle{NPE for LSST}
\shortauthors{Yicun Duan, Xinyue Li, Camille Avestruz, Jeffrey Regier and LSST Dark Energy Science Collaboration}
\graphicspath{{./}{figures/}}

\begin{document}

\title{Neural Posterior Estimation for Cataloging Astronomical Images from the Legacy Survey of Space and Time}

\author{Yicun Duan}
\affiliation{Department of Statistics, University of Michigan}

\author{Xinyue Li} 
\affiliation{Department of Statistics, University of Michigan}

\author[0000-0001-8868-0810]{Camille Avestruz}
\affiliation{Department of Physics, University of Michigan}

\author[0000-0002-1472-5235]{Jeffrey Regier}
\affiliation{Department of Statistics, University of Michigan}
\email{cavestru@umich.edu, regier@umich.edu}

\author{LSST Dark Energy Science Collaboration}
\noaffiliation



\begin{abstract}

The Vera C. Rubin Observatory Legacy Survey of Space and Time (LSST) will commence full-scale operations in 2026, yielding an unprecedented volume of astronomical images. Constructing an astronomical catalog, a table of imaged stars, galaxies, and their properties, is a fundamental step in most scientific workflows based on astronomical image data. Traditional deterministic cataloging methods lack statistical coherence as cataloging is an ill-posed problem, while existing probabilistic approaches suffer from computational inefficiency, inaccuracy, or the inability to perform inference with multiband coadded images, the primary output format for LSST images. In this article,  we explore a recently developed Bayesian inference method called neural posterior estimation (NPE) as an approach to cataloging. NPE leverages deep learning to achieve both computational efficiency and high accuracy. When evaluated on the DC2 Simulated Sky Survey---a highly realistic synthetic dataset designed to mimic LSST data---NPE systematically outperforms the standard LSST pipeline in light source detection, flux measurement, star/galaxy classification, and galaxy shape measurement. Additionally, NPE provides well-calibrated posterior approximations. These promising results, obtained using simulated data, illustrate the potential of NPE in the absence of model misspecification. Although some degree of model misspecification is inevitable in the application of NPE to real LSST images, there are a variety of strategies to mitigate its effects.

\end{abstract}



\section{Introduction}

The Vera C. Rubin Observatory Legacy Survey of Space and Time (LSST) will begin full-scale operation in 2026.
During its 10-year lifespan, LSST will collect 500 petabytes of images of the night sky in the southern hemisphere. The survey aims to address questions about the structure and evolution of the Universe pertaining to the nature of dark energy and dark matter, the contents of the solar system and the transient optical sky, and the structure of the Milky Way \citep{Ivezić_2019}.

Most scientific inquiries based on LSST data will not use the survey images directly; researchers will instead rely on astronomical catalogs constructed from LSST images. Astronomical catalogs are tables of light sources (a.k.a. objects) and their properties, such as position, source type (e.g., star or galaxy), flux (brightness), and shape. Catalog construction is challenging because it requires solving an ill-posed inverse problem since multiple catalogs can be consistent with the data.
Because the majority of imaged light sources are faint in terms of signal-to-noise ratio, the detection of even isolated light sources is prone to error.

Inferring the correct catalog for imaged light sources that visually overlap, known as blended objects or blends, is even more challenging; it requires separating each light source from both background noise and other light sources.
As advances in telescope technology improve light sensitivity, resulting in a higher density of light sources in the images captured, blending has become increasingly common \citep{melchior2021challenge}.
LSST will observe a much higher density of sources than previous wide-field surveys.
Projections indicate that 62\% of the LSST-imaged galaxies will be blended to some extent \citep{sanchez2021effects}, which is in line with figures for the Hyper Suprime-Cam survey, in which 58\% of objects are blended \citep{bosch2017hyper}.

Traditional algorithmic approaches to cataloging, such as the approaches of SExtractor \citep{bertin1996sextractor} and the planned LSST image processing pipeline \citep{bosch2018overview}, do not address the ambiguity posed by the increased volume and enhanced resolution of LSST images in a statistically coherent way.
For instance, in the LSST pipeline, detections are made independently for each image band using a peak-finding routine. Following this, detections across bands are merged according to a matching criterion.
Blended objects are often subject to over-merging during this stage, so candidate blends are subsequently processed by a deblending algorithm.
The primary deblending algorithm employed by LSST, known as Scarlet, operates under the assumption that light sources are compact and their spectra are spatially uniform \citep{melchior2018scarlet}.
Despite the inherent ambiguity in detecting and characterizing light sources, traditional algorithms ultimately produce a single realization of the catalog.

Probabilistic cataloging is an alternative to traditional algorithmic approaches \citep{brewer2013probabilistic, portillo2017improved, feder2020multiband, buchanan2023markov}.
In probabilistic cataloging, a distribution of latent catalogs is inferred conditional on the images, which are treated as observed random variables in a generative model.
Deblending is a consequence of performing posterior inference, eliminating the need for special process logic to address potential blends.
In addition, probabilistic cataloging effectively handles multiband imaging, in which each image captures photons in different frequency ranges, by modeling each image as a realization of the same latent catalog \citep{feder2020multiband}.

Unfortunately, existing probabilistic cataloging methods cannot be applied to the complete LSST dataset. Probabilistic cataloging methods based on Markov chain Monte Carlo (MCMC), including \citet{brewer2013probabilistic, portillo2017improved, feder2020multiband} and \citet{buchanan2023markov}, are not computationally efficient enough to apply to the entire collection of LSST images; these methods are designed to process small cutouts of much larger images.
Additionally, \citet{brewer2013probabilistic, portillo2017improved, feder2020multiband} and \citet{liu2023variational} are limited to detecting only stars, rather than both stars and galaxies.
\citet{regier2019approximate} refines the estimates of preexisting catalogs, but is unable to identify new stars and galaxies or account for detection ambiguity.

Further, no previous probabilistic methods are compatible with coadded astronomical images even though coadded images will be the primary format used to distribute LSST image data.
Coadded images are formed by merging multiple single-exposure images of the same area of the sky, creating deeper images with improved signal-to-noise ratio.
One key technical advantage of coadded images is their significantly reduced data size compared to the collective size of the single exposures from which they are derived. For LSST, the complete collection of coadded images will amount to approximately 75 TB, while the raw exposures will exceed 15 PB in size. 

Because raw exposure images are not pixel-aligned, combining them requires resampling to establish a common grid. This resampling process introduces correlation among neighboring pixel intensities because each resampled pixel value is a weighted average of overlapping sets of raw pixel intensities and, more subtly, because the alignment of images to a common grid is determined by the raw pixel intensities. Whereas the raw pixel intensities can plausibly be considered conditionally independent, the resampled pixel intensities are coupled.
The correct generative model of coadded images includes a latent variable for each pixel of each single exposure, which amounts to more than a quadrillion latent variables for LSST.
These latent variables are not easily marginalized over in MCMC or traditional variational inference procedures.

In this work, we explore an approach to Bayesian inference known as neural posterior estimation (NPE).
NPE is a type of likelihood-free amortized variational inference that fits a variational distribution to minimize an expectation of the Kullback-Leibler (KL) divergence from the posterior to the variational distribution \citep{zhang2021real}.
Algorithmically, NPE amounts to simulating data (catalogs and images) from the generative model and fitting a neural network (the inference network) to predict the corresponding catalog.
Recently, NPE has been shown to be globally convergent, with plausible assumptions, despite the non-convexity of neural networks with respect to their parameters \citep{mcnamara2024globally}.
Most importantly, for coadded images, NPE implicitly marginalizes nuisance latent variables \citep{ambrogioni2019forward}.
Thus, despite performing inference with a properly specified Bayesian model that incorporates an intractable quantity of nuisance latent variables, NPE remains computationally efficient.

We assess the potential for cataloging LSST with NPE by applying NPE to the LSST-DESC DC2 Simulated Sky Survey (\cref{sec:model}).
This simulated dataset, created in preparation for the arrival of the LSST images, was designed by the LSST Dark Energy Science Collaboration (LSST DESC) to be highly realistic.
As a simulated dataset, DC2 includes a ground truth catalog corresponding to each image, which is invaluable for validation purposes.
This simulated data also effectively defines our statistical model: because NPE is a likelihood-free inference method, it does not require the generative model to be explicitly specified.
Instead, it suffices that complete data (i.e., latent catalog and observed images) sampled from it are available for fitting the inference network, which defines the variational distribution.

In our collaboration's previous work with NPE for probabilistic cataloging \citep{liu2023variational,hansen2022scalable,patel2025neural,regier2025neural}, we performed inference exclusively with single-exposure images, and in a workshop paper, we put forward the idea of using NPE for coadd images \citep{wang2022statistical}. In this work, we apply an NPE method (\cref{sec:methodology}) to images that are coadded and multiband. To the best of our knowledge, this is the first archival publication to report applying Bayesian inference to coadded astronomical images with a model that accounts for the dependence among pixels due to the coaddition process.

There is not an established methodology for benchmarking probabilistic approaches to astronomical cataloging, so we develop one (\cref{sec:eval_strategy}). We compare the proposed method to the LSST pipeline, making use of the  availability of ground-truth catalogs in the DC2 dataset.

Our results show that the proposed method outperforms the LSST pipeline in detection performance, flux estimation, star/galaxy classification, and galaxy ellipticity estimation (\cref{experiments}).
NPE outperforms the LSST pipeline at all tasks, across all magnitudes and levels of blending.
The calibration of our posterior approximations is good but not perfect.
There are many ways to potentially improve the calibration by making our variational family more expressive.

Throughout our experiments, we used training and test data drawn from the same distribution (DC2).
However, in the future, when we apply our software to real LSST images, the training set will still be made up of simulated data and therefore at least somewhat misspecified (\cref{sec:discussion}).
We suggest approaches to mitigating the effects of model misspecification through a combination of domain adaptation techniques and modeling enhancements that can be explored in future work.
With a more flexible model, inference could be performed using the expectation-maximization (EM) algorithm, which, under certain conditions, can be a statistically consistent estimation procedure. Within this EM procedure, the proposed NPE method could serve as a powerful solver for each E-step (\cref{nonparametric}).

\section{The LSST DESC DC2 Simulated Sky Survey}
\label{sec:model}
In preparation for the arrival of LSST images, the LSST Dark Energy Science Collaboration created the DC2 dataset as a testbed for image-processing software currently in development \citep{korytov2019cosmodc2, abolfathi2021lsst, abolfathi2021desc}. The dataset has three components: a large synthetic galaxy catalog designed for dark energy science (\cref{sec:cosmodc2}); a collection of simulated unprocessed images, resembling the images read off a CCD (\cref{sec:imsim}); and a collection of calibrated, coadded, and otherwise processed images (\cref{sec:coaddition}). 
The DC2 simulator effectively serves as the generative model for the proposed method (\cref{sec:dc2-bayesian}) as well as providing our validation data.

\subsection{Catalog simulation}
\label{sec:cosmodc2}

CosmoDC2, introduced by \cite{korytov2019cosmodc2}, is a sophisticated synthetic galaxy catalog developed to aid dark energy science analyses, specifically aimed at constraining cosmological parameters.
The CosmoDC2 simulation incorporates three inputs:
\begin{enumerate}[topsep=1pt,itemsep=0pt]
    \item The Outer Rim simulation \citep{heitmann2019}, a trillion-particle n-body simulation. Outer Rim is one of the largest simulations at its mass resolution and was run using the Hardware/Hybrid Accelerated Cosmology Code \citep{habib2016hacc}.
    \item The AlphaQ simulation, which covers a volume 1600 times smaller than the Outer Rim simulation but provides more complex galaxy properties.
    \item The Universe Machine catalog \citep{behroozi2019universemachine}, which contains the output of an n-body simulation that models the connection between galaxies and dark matter halos. In the simulation, galaxy positions, stellar masses, and star formation rates are empirically calibrated with observational data.
\end{enumerate}

\vspace{2pt}
\noindent From these inputs, three intermediate products are generated:
\begin{enumerate}[topsep=1pt,itemsep=0pt]
    \item The BaseDC2 catalog is formed using halo and merger-tree information from the Outer Rim simulation combined with resampling of galaxies from the Universe Machine catalog.
    \item The lensing maps are generated by applying a ray-tracing code to light cones based on Outer Rim simulation particle data. Here, particles are projected onto redshift shells, from which one can compute the surface density.  The path of photons through these shells produces curved-sky lensing maps that encode photon deflections and the distortion of galaxies due to weak gravitational lensing.
    \item The Galacticus library is based on a semi-analytic model of galaxy formation \citep{benson2012galacticus} applied to the AlphaQ simulation.  The semi-analytic model maps galaxy evolution to the dark matter halo mergers of the simulation.
\end{enumerate}

\vspace{2pt}
To create the CosmoDC2 catalog, galaxies from the BaseDC2 catalog are matched, based on their fluxes, with those in the Galacticus library and are paired with lensing information derived from the lensing maps.
Subsequently, ultra-faint galaxies that are disconnected from resolved halos are included in the catalog, and the impact of intergalactic dust is determined through ray tracing.

Finally, to create the DC2 truth table (see Appendix~A), the CosmoDC2 galaxy catalog is augmented with stars and supernovae.

\subsection{Simulation of raw images}
\label{sec:imsim}

The DC2 simulator takes the DC2 truth table (described above) and produces images covering an area of approximately 300 deg$^2$, which is large enough for studying weak lensing and large-scale structure.
The simulated visit strategy (that is, the sequence of regions imaged) reflects realistic cadences for the first 5 years of LSST operation.
The visits include both translational and rotational dithers, so that, for example, a particular star does not always appear in the same within-pixel position across exposures.

The simulated images include six optical bands, each composed of $4000\times4000$ pixels.
The images are simulated using GalSim \citep{rowe2015galsim} in conjunction with LSST-specific GalSim extensions provided by the imSim package. The imSim package, in turn, is based on the Rubin Observatory image simulation package, which provides a simulator of the LSST camera (LSSTCam) that models its 189 science charge-coupled devices (CCDs), each with 16 megapixels.
The imSim package models the characteristics of the LSST filter, the brightness of the sky (background), and image artifacts, such as cosmic rays, pixel saturation, and bleed trails.

imSim also provides a model of the point spread function (PSF) based on a frozen-flow approximation for six atmospheric turbulent layers with plausible heights and outer scales, optical aberrations (based on actual optic corrections from LSSTCam), chromatic dependence, and the brighter-fatter effect.

\subsection{Image processing and coaddition}
\label{sec:coaddition}
\begin{figure*}
    \centering
    \includegraphics[width=0.48\textwidth]{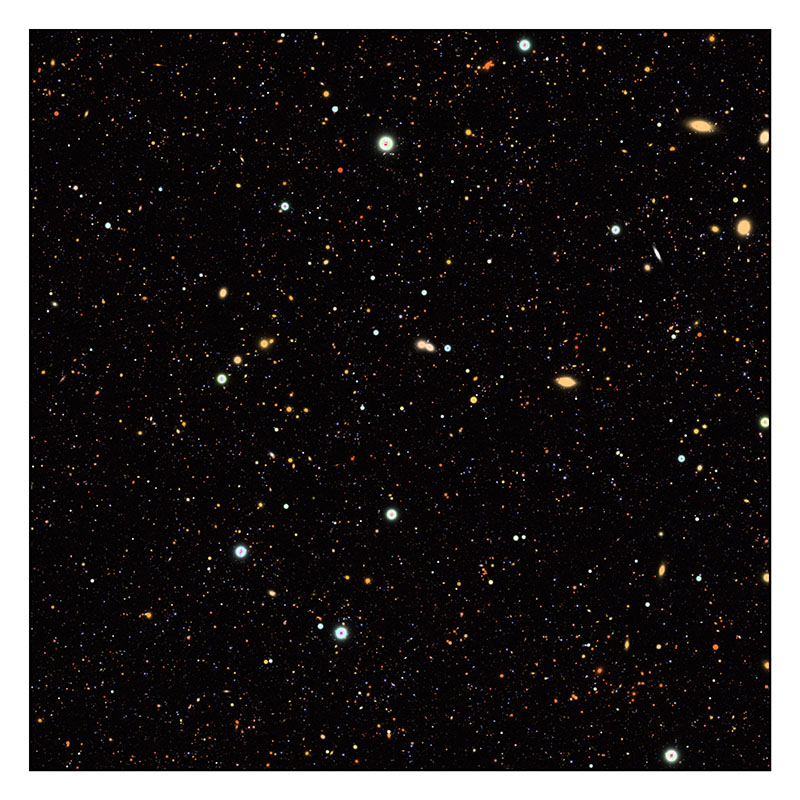}
    \includegraphics[width=0.48\textwidth]{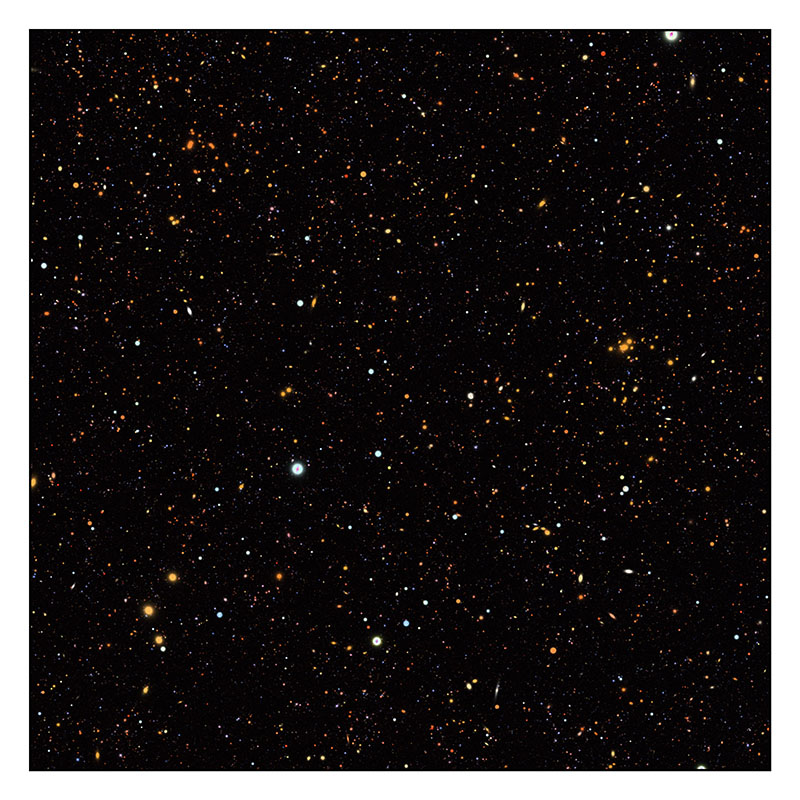}
    \caption{Coadded DC2 images for two DC2 patches. Each image is 4100$\times$4100 pixel. Each patch contains tens of thousands light sources. Hundreds of them are bright enough to be visible in these renderings.}
    \label{fig:dc2_images}
\end{figure*}

In DC2, the simulated raw images are processed by the same Data Release Production (DRP) pipeline that will be applied to the real raw LSST images.

The first step in processing these images, known as image signature removal, produces a version of the image that is more homogeneous across instruments.
Image signature removal includes bias subtraction, cross-talk correction, non-linearity correction, flat fielding, brighter-fatter correction, and masking of bad and saturated pixels.

The next step is image characterization, which includes background estimation and subtraction, PSF modeling,
cosmic-ray detection and repair, source detection, source deblending, and source measurement. 
Once the fluxes of prominent imaged sources have been estimated in terms of image-specific photoelectron counts, the entire image can be re-expressed based on an external reference catalog that lists the flux of these sources in physical units.
The resulting images are known as ``calibrated exposures.''

In DC2 (as in LSST), the surveyed area is divided into square tracts, and the tracts, in turn, are divided into a 7$\times$7 grid of patches.
For each patch, a coadded image is generated from the calibrated exposures that overlap with the patch.
The overlapping calibrated exposures are first resampled to a common pixelization and then combined with a weighted mean.
The resulting coadded images are each 4100$\times$4100 pixels with a pixel scale of 0.2 arc seconds.
\cref{fig:dc2_images} provides examples of these images.

\subsection{DC2 as a Bayesian model}
\label{sec:dc2-bayesian}
The DC2 simulator implicitly specifies a Bayesian model as it defines a generative process over both latent quantities (i.e., astronomical object properties, pixel-level shot and read noise) and coadd images given these quantities.
The prior in this model depends on a fixed cosmological realization provided by the three inputs to the CosmoDC2 simulation: the Outer Rim simulation, the AlphaQ simulation, and the UniverseMachine catalog.
The prior is a stochastic model for assigning galaxy properties to halos; samples of it form the base DC2 catalog.
One may additionally think of the pixel-level shot and read noise as part of the prior, as it is unobserved nuisance variation. From this perspective, the conditional likelihood of the coadded images, which are deterministically computed from the raw individual exposures, would be a delta function.
Alternatively, we may think of the conditional likelihood as marginalized over pixel-level noise, leaving just the catalog-level stochasticity as the prior.

\section{Methodology: Bayesian inference}
\label{sec:methodology}

We seek to infer a particularly important marginal of the posterior distribution (\cref{target}).
To do so, we use variational inference with a spatially autoregressive family of amortized variational distributions (\cref{variational_family}).
To fit the variational distribution, we use a type of variational inference called neural posterior estimation (NPE), which implicitly marginalizes over nuisance variables, has global convergence guarantees, and is likelihood-free (\cref{npe}).

The method development in this section follows that of \cite{regier2025neural}, which introduces a spatially autoregressive variational distribution for the detection of astronomical objects, and establishes its theoretical and empirical properties.
Here, we generalize this variational distribution to apply to multiband coadded images.

\subsection{Our target: a marginal of the posterior}
\label{target}

Let $x$ represent a coadded multiband image that is $H \times W$ pixels in size.
Let $\theta_0$ represent the catalog of astronomical objects whose centroids are contained in the region imaged by $x$.
We treat $x$ as an observed variable and $\theta$ as a latent random variable.

We restrict our attention to the subset of sources $\theta_1 \subset \theta_0$ whose flux exceeds a user-defined threshold.
In DC2 (and essentially any survey), the vast majority of light sources are below the detection limit.
Filtering the light source by flux focuses our inference procedure on light sources that can be detected with useful accuracy, rather than simply recovering the prior.

Next, we re-express the catalog $\theta_1$ in a form that reflects the spatial structure of an image, as a spatial grid of ``tiles.''
Let $T$ be the side length of the tiles, expressed in the coordinates of the image. We set $T$ to 4, which implies that tiles are $4\times{}4$-pixel regions.
Let $H' := H / T$ and $W' := W / T$.
For $h=1,\ldots,H'$ and $w=1,\ldots,W'$, let $\theta_2^{(h,w)} \subset \theta_1$ contain the sources whose positions, expressed in pixel coordinates, are in $[Th,Tw)\times[T(h+1), T(w+1))$.

Then, let $z_{h,w} \subset \theta_2^{(h,w)}$ denote the singleton set containing the source with the greatest flux in $z_{h,w}$ if $\theta_2^{(h,w)}$ is nonempty, or else the empty set.
This change of variables is motivated by the need to express the catalog as a bounded number of sources in each region, allowing at most one source in each region of the image.
 
Finally, let $z := \cup_{h,w} z_{h,w}$. We seek to infer $p(z \mid x)$.

\subsection{Our variational family}
\label{variational_family}
To approximate $p(z \mid x)$, we use a spatially autoregressive variational distribution that factorizes according to a partial ordering of the tiles.
To define this partial ordering, set the number of ranks $K = 4$ and let the rank of the tile with index $(h, w)$ be
\begin{align*}
    \Psi(h, w) = (h \bmod \sqrt{K}) \cdot \sqrt{K} + (w \bmod \sqrt{K}).
\end{align*}
This scheme ensures that neighboring tiles (adjacent or diagonally adjacent) have distinct ranks, which is important for detecting light sources whose positions are near the borders of tiles.
Let $\Omega := \{1, \ldots, H'\} \times \{1, \ldots, W'\}$ denote the indices of all the tiles, and for $k=1,\ldots,K$, let
\begin{align*}
        A_k := \{ (h, w) \in \Omega : \Psi(h, w) = k \}
\end{align*}
denote the set of tiles with rank $k$.
For $(h, w) \in \Omega$ and $\ell = (h,w)$, let $z_\ell := z_{h,w}$ serve as shorthand for the latent variables corresponding to tiles with index $(h, w)$.
For $k=1,\ldots,K$, denote the pre-rank-$k$ latent variables as $z_{<k} := \{z_\ell : \ell \in \cup_{k' < k} A_{k'}\}$.

To approximate $p(z \mid x)$, we use a variational distribution with the following form:
\begin{align*}
    q(z \mid x) := \prod_{k = 1}^K \prod_{(h, w) \in A_k} q(z_{h,w} \mid z_{<k}, x).
\end{align*}
\cite{regier2025neural} establishes conditions under which a variational distribution of this form mirrors the conditional independence assertions of the target distribution $p(z \mid x)$, in the sense of being a ``minimally sufficient I-map.'' This is a desirable property for a variational distribution because it implies that the variational distribution has the flexibility to model dependencies in the target distribution, but not enough flexibility to model dependencies that are known a priori to be absent in the target distribution.

We refer to $q(z_{h,w} \mid z_{<k}, x)$ as the per-tile variational distribution.
The probabilistic structure of the per-tile variational distribution is given in \cref{vardisttable}; for each tile, this distribution is governed by 22 distribution parameters.

For a tile with index $(h, w)$, these distributional parameters are outputted in spatial position $(h, w)$ by a fully convolutional neural network that takes $z_{<k}$ and $x$ as input.
This network is itself composed of three fully convolutional networks.
The first, called the neighborhood network, takes as input the partially observed catalog $z_{<k}$, represented by a fixed-dimensional $H' \times W' \times D$ tensor, where $D$ is the number of parameters conditioned on and the tiles whose indices have rank greater than or equal to $k$ have been masked.
This network outputs a representation of the neighborhood for each tile.
The second network, called the image backbone, takes an $H \times W \times C$ as input, where $C = 6$ is the number of filter bands, and yields an $H' \times W' \times 256$-dimensional representation.
The third, called the detection network, combines the outputs of the other two networks to produce an $H' \times W' \times 22$-tensor of unconstrained distributional parameters, which are then rescaled as needed (e.g., to be strictly positive) to parameterize the factors of the per-tile variational distribution in \cref{vardisttable}.
The weights of these convolutional networks, collectively denoted as $\phi$, are the variational parameters. Appendix~B provides additional details of the BLISS neural network architecture.

\begin{table*}
    \centering
    \begin{tabular}{@{}llll@{}}
        \toprule
        Latent variable & Distribution & \# Parameters & Conditional on\\
        \cmidrule{1-4}
        Has a source & Bernoulli & 1 & -- \\
        Position & Truncated Bivariate Normal & 4 & Has a source = True\\
        Source type & Bernoulli & 1 & Has a source = True \\
        Flux & Multivariate Log Normal & 12 & Has a source = True \\
        Ellipticity & Bivariate Normal & 4 & Source type = Galaxy \\
        \bottomrule
    \end{tabular}
    \caption{Factors of the per-tile variational distribution. }
    \label{vardisttable}
\end{table*}

\subsection{Neural Posterior Estimation}
\label{npe}

To make explicit the dependence of the variational distribution on $\phi$, we set $q_\phi(z \mid x) := q(z \mid x)$.
In NPE, the variational distribution is fitted to minimize the following expectation of the forward KL divergence:
\begin{align*}
    \mathcal L := \mathbb E_{x \sim p(x)} \left[ \KL{p(z \mid x)}{q_\phi(z \mid x)} \right]. 
\end{align*}
The expectation here is taken with respect to synthetic images sampled from the generative model.
The gradient of the variational objective is
\begin{align*}
    \nabla \mathcal L = -\mathbb E_{x, z \sim p(x, z)} \left[ \nabla q_\phi(z \mid x) \right].
\end{align*}
Unbiased approximations of this gradient are available through Monte Carlo sampling.

The gradient of the objective can be further expanded by substituting in our particular variational distribution:
\begin{align*}
\nabla \mathcal L(\phi)
&= -\mathbb{E}_{x,z\sim p(x, z)} \left[\sum_{k = 1}^K \sum_{\ell \in C_k} \nabla \log q_\phi(z_\ell \mid z_{<k}, x) \right].
\end{align*}

These stochastic gradients are used to iteratively minimize the objective. Once $\phi$ is fitted, the observed data $x_0$ is supplied to the inference network, which yields the posterior approximation $q_\phi(z \mid x_0)$.

Minimizing this objective can be viewed as learning to probabilistically infill catalogs: given a subset of a catalog, $z_{<k}$, infer catalog $z_{k}$ (in the context of image $x$).
The ordering of the ranks, although necessary to define a valid variational distribution that can be sampled, is largely arbitrary.
Thus, in practice, our approach to minimizing the objective is to sample random tile masks and to train our inference network to infer the unmasked parts of the catalog.
Randomizing tile ranks across training iterations provides regularization by requiring the neural network to deal with arbitrary masking patterns.

During training, we additionally regularize by performing on-the-fly data augmentation.
We follow standard practice in neural network performance optimization, which involves using validation loss to search for good network architectures and tuning parameters \citep{karpathy2019recipe}.
Training is computationally intensive, requiring 22 hours on a single GPU.
Inference is extremely fast, with data loading (moving images from disk to CPU memory) acting as the bottleneck rather than the neural network forward passes required to process that data.
Appendix~B provides further details of the proposed algorithm, including illustrations of the neural network architecture, projected runtimes for our methods on the complete LSST coadd image set, and discussion about storing probabilistic LSST catalogs.

\section{Performance evaluation strategy}
\label{sec:eval_strategy}
In \cref{experiments}, our experimental results are grouped by task: detection, flux estimation, star/galaxy classification, and galaxy shape parameter estimation. In this section, we discuss cross-cutting aspects of our performance evaluation strategy.

For each task, we perform two complementary types of performance evaluation.
In the first, we assess the performance of our posterior approximation in terms of a single point estimate, which we set to be the median of the variational distribution.
This approach has advantages and drawbacks.
The chief advantage of this approach is that it lets us directly compare the performance of our probabilistic method to traditional (non-probabilistic) methods, which typically only output point estimates.
In particular, we compare our method to the LSST coadd catalog (\cref{lsst-catalog}), which serves as our primary reference point for judging performance.
To evaluate point estimates, a bipartite match is formed between detections and true objects (\cref{matching}).
A bootstrapping procedure (\cref{sec:bootstrap}) assesses the stability of these results.

To complement point-estimate-based metrics, which do not consider the calibration of the posterior approximation, we consider two measures of posterior calibration: the frequentist coverage of credible intervals (\cref{credible-intervals}) and calibration plots (\cref{calibration-plots}).

We typically stratify the metrics based on true flux (\cref{flux_units},\cref{flux_cut}) or blendedness of objects (\cref{blending_units}). The results we report are all based on a fairly constructed test set (\cref{held-out}).
We give more details about each of these aspects of our performance evaluation strategy in the following subsections.

\subsection{Our primary benchmark: the LSST coadd catalog}
\label{lsst-catalog}

The LSST pipeline constructs catalogs of objects in coadded images through a sequential multi-step process \citep{bosch2017hyper}.

First, source detection is performed independently in each band using a maximum-likelihood detection algorithm with a 5$\sigma$ threshold. This involves convolving the image with a Gaussian filter matched to the size of the coadded PSF, identifying regions (footprints) that exceed the threshold. Each footprint may contain multiple peaks, each representing a potential source.

The footprints and their peaks are then merged across different bands, processed in a priority order based on signal-to-noise ratios. Overlapping footprints are combined, and peaks are matched or added based on their proximity to previously identified peaks. This merging step helps to create a catalog of consistent footprints and peaks across all bands while attempting to cull spurious detections.

Following the merging, deblending is carried out independently in each band to separate blended sources within each footprint, yielding individual measurements for each detected source. This step ensures that the catalog contains the same objects in every band, although deblending is performed independently per band.

Each object is then assigned a reference band chosen based on its detection significance and signal-to-noise ratios in various bands. This reference band provides the most reliable measurements for the object.

Finally, forced measurements are performed across all bands using the position and shape parameters of the reference band. This ensures consistent measurements and reliable computation of object colors by accounting for differences in the effective PSFs of different bands.

\subsection{Matching detections and true objects}
\label{matching}
To evaluate the accuracy of a catalog, we first match cataloged detections and true objects.
The true objects available for matching are those exceeding 50 Jy in flux.
For each of these objects, we identify the corresponding detection by selecting the nearest detected object within a one-pixel radius.
A one-pixel radius is within the range of matching thresholds used in other publications: \citet{sanchez2020data} reports using a three-pixel radius for DC1 data and \citet{portillo2017improved} reports using a half-pixel radius for SDSS data. 
Appendix~C considers additional matching thresholds, namely three and five pixels; in short, the general trends in relative performance are not changed by the precise threshold used.
Among detection-object pairs within the specified radius, we then find the optimal (distance-minimizing) bipartite matching using the Hungarian algorithm.
The DC2 catalog provides a matching based on a different matching algorithm, but we exclusively use our matching procedure to ensure that the matching is done in the same way for both BLISS and the LSST pipeline.

\subsection{Bootstrap confidence intervals}
\label{sec:bootstrap}

In some performance plots, we represent performance variability with a shaded region. The width of this shaded region is two standard deviations, which we compute using the nonparametric bootstrap.
The bootstrap datasets are formed by sampling 80$\times$80-pixel sub-images (and the corresponding ground-truth catalogs) from the test set with replacement. 
As such, the width of the shaded region does not account for performance variability due to the composition of the training set: the variability it reflects is solely from the draw of the test set.
Sometimes, the performance of two methods and corresponding shaded regions are shown on the same plot. Overlapping of the shaded regions does not mean that the difference between the two methods is statistically insignificant: Because both methods are tested on the same test data (i.e., a ``matched pairs'' experiment), the combined width of these shaded regions overstates the variability in the difference in their performances.

\subsection{Frequentist coverage of credible intervals}
\label{credible-intervals}
In addition to evaluating the accuracy of point estimates based on the variational distribution, we evaluate the calibration of the variational distribution by forming credible intervals based on it.
In particular, for the variational distribution of each image in the test set, we form highest-density credible intervals for various marginals, targeting various coverage levels.
We then check how frequently the credible intervals include the ground truth.
If the variational distribution is a well-calibrated approximation of the posterior, the nominal coverage level will be similar to the empirical/frequentist coverage of the credible intervals.
If the nominal coverage of the credible intervals exceeds the empirical coverage, the variational distribution is likely underdispersed; if the empirical coverage exceeds the nominal coverage, the variational distribution is likely overdispersed.

Agreement of the nominal and empirical coverage levels is necessary but not sufficient to ensure calibration. To see this, observe that if the variational distribution were set to the prior, empirical and nominal coverage levels would agree perfectly.
Next, we present a complementary probe of calibration that is not ``gamed'' by setting the variational distribution to the prior.

\subsection{Posterior calibration plots}
\label{calibration-plots}

Calibration plots are graphical tools for assessing the performance of probabilistic predictive models by comparing predicted probabilities with actual outcomes. A calibration plot includes a 45-degree line, often referred to as the line of perfect calibration, which represents the ideal scenario where predicted probabilities match the true probabilities.

Our calibration plots are based on simulated data, so the ``actual outcome'' is the ground truth, which appears on the x-axis.
As our estimates, we use 90\% credible intervals, which we display as vertical line segments (an interval plot).
Ideally, the credible intervals would be narrow while nevertheless achieving 90\% coverage of the 45-degree line.

\subsection{Units of flux}
\label{flux_units}
The metrics we report are typically stratified by flux because performance depends so heavily on it: brighter sources are easier to characterize, while most sources are so faint as to be undetectable.

The flux (or brightness) of an object directly influences the accuracy with which a method can detect that object.  We quantify the brightness of an object in units of AB magnitude \citep{lsst2018choice_of_flux_unit}. In this system, magnitude is defined as $-2.5 \log_{10}(F_b / F_{AB})$, where $F_b$ is the flux of an object in Jy (Jansky) and $F_{AB}$ is the zero point, which is set to 3631 Jy. Note that smaller magnitudes correspond to brighter sources.

By default, we stratify and/or cut based on magnitude in the r band, which often has the highest signal-to-noise ratio.

\subsection{Flux cuts}
\label{flux_cut}
The majority (85.13\%) of light sources in DC2 are fainter than 26.4 magnitude in the r band, which is so faint as to be nearly undetectable. It can be seen that 26.4 magnitude is close to the detection limit in two ways. First, the right-most bin in \cref{fig:detection_performance} shows that the LSST pipeline only detects 55\% of sources fainter than 25.6 magnitude and that half of its detections in this range are false detections. Second, 26.9 magnitude in the r-band is projected to be the 5-sigma point source depth for LSST coadds \citep{RubinKeyNumbers}, which are based on twice as many images (10 years rather than 5 years) as the DC2 images.

During training, we filter out these light sources fainter than 26.4 magnitude in the r band from the catalogs. In the notation of our generative model, filtering sources amounts to designating them as elements of $\theta_1 \setminus \theta_0$.
For training, we additionally filter out 0.89\% of the remaining light sources so that no tile contains more than one light source.
These latter filtered-out light sources, though not used during training, nevertheless count against our method during performance evaluation. The restriction of one light source per tile limits our ability to detect light sources whose centroids are less than 4 pixels apart.  However, extending our methodology to two light sources per tile does not improve detection performance.

\subsection{Units of blending}
\label{blending_units}
We report some metrics stratified by blendedness, typically in the context of a minimum flux threshold.
We adopt the high-level approach to quantifying blendedness of \citet{bosch2017hyper}, in which blendedness is
$1 - \frac{I_{\textrm{child}}}{I_{\textrm{parent}}}$,
where $I_{\textrm{child}}$ is the flux of the object in question (true or predicted, depending on context) and $I_{\textrm{parent}}$ is the overall flux in the vicinity of this object.
The closer this value is to 1, the more blended the object is. 

We define $I_\textrm{child}$ and $I_\textrm{parent}$ slightly differently than \citet{bosch2017hyper} because we alone have access to ground-truth catalogs by virtue of working with synthetic data. We take $I_\textrm{child}$ to be the true flux of the object in question (that is, without shot noise) and 
we take $I_{\textrm{parent}}$ to be the Gaussian-weighted sum of the true fluxes of neighboring objects. The covariance is isotropic with a standard deviation of 6 pixels (1.2 arc seconds), and we truncate beyond 30 pixels (6 arc seconds).

\subsection{Held-out data}
\label{held-out}
We partition the DC2 data into training, validation, and test sets at the patch level, prior to dividing the patches into $80 \times 80$-pixel sub-images. This limits information leakage due to spatial correlation and provides a more realistic assessment of how our method would perform on new patches. Of the 98 DC2 patches for which coadd images are publicly available, we use 78 for training, 10 for validation, and 10 for testing. The validation loss is used to decide on early stopping, which provides regularization, and for tuning our network architecture. All results reported in Section~\ref{experiments} describe inference on the test set.

\section{Experimental Results}
\label{experiments}
In this section, we refer to our implementation of the proposed NPE method as the Bayesian Light Source Separator (BLISS).
BLISS is a long-running software project \citep{liu2023variational,hansen2022scalable,patel2025neural,regier2025neural}, which we have extended in this work to apply to the DC2 dataset.
Our experiments use the benchmark methodology from \cref{sec:eval_strategy} to assess the performance of BLISS in object detection, flux estimation, star/galaxy classification, and galaxy shape estimation.

\subsection{Detection}

\begin{figure*}
    \centering
    \subfigure(a){\includegraphics[width=0.45\textwidth]{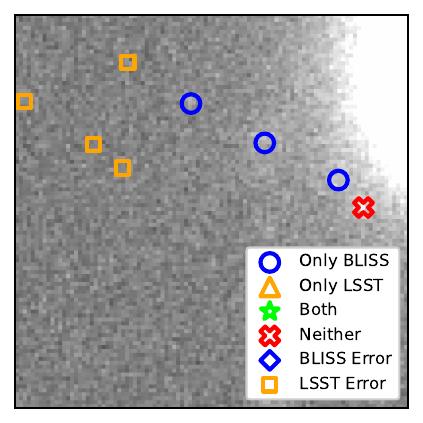}} 
    \subfigure(b){\includegraphics[width=0.45\textwidth]{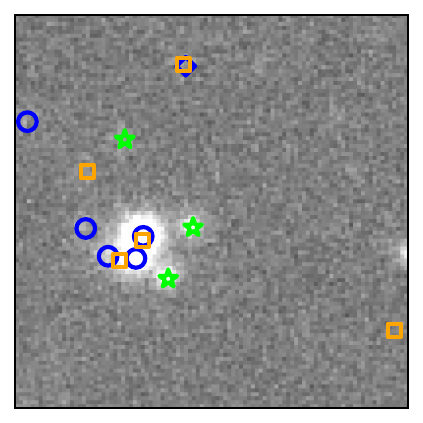}} 
    \subfigure(c){\includegraphics[width=0.45\textwidth]{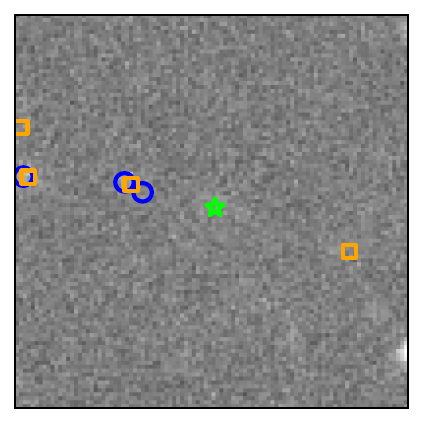}}
    \subfigure(d){\includegraphics[width=0.45\textwidth]{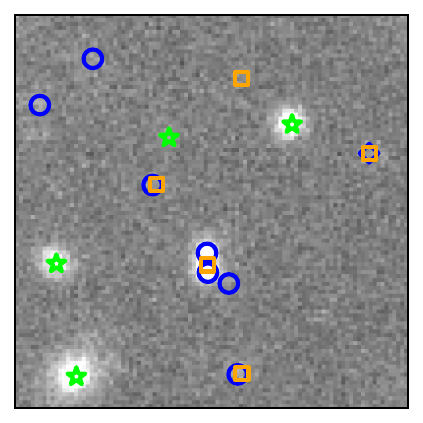}}
    \caption{Examples of BLISS and LSST detections. BLISS detects some faint and/or overlapping sources missed by the LSST detection pipeline. The legend marks sources detected only by BLISS (circles), only by LSST (triangles), by both (stars), by neither (crosses), with BLISS localization errors (diamonds), and with LSST localization errors (squares). Several LSST predictions may appear to be correct, but they are in fact off by more than our one-pixel threshold. These images are $80 \times 80$ pixels captured in the r band.
    }
    \label{fig:detection_example}
\end{figure*}

\cref{fig:detection_example} shows examples of detections in four regions of DC2.  The circles indicate objects that only BLISS detected, the triangles indicate objects that only the LSST pipeline detected, and the stars indicate objects that both BLISS and the LSST pipeline detected.
Light sources that are both bright and isolated, such as those in the top-right and bottom-right panels, tend to be detected by both methods.
Most light sources detected by the LSST pipeline were also detected by BLISS, so no triangles appear in these examples.
Several light sources that are not easily visible by eye were successfully detected by BLISS but not the LSST pipeline. The diamonds and squares indicate false detections by either method. False detections often occur in situations where the background could be mistaken for an object.

\begin{figure*}[t]
    \includegraphics[width=0.47\textwidth]{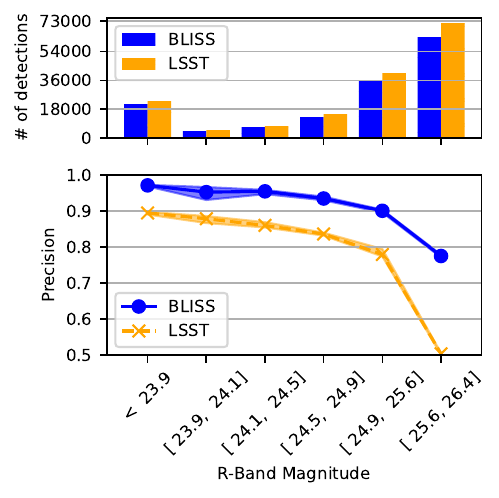}
    \hfill
    \includegraphics[width=0.47\textwidth]{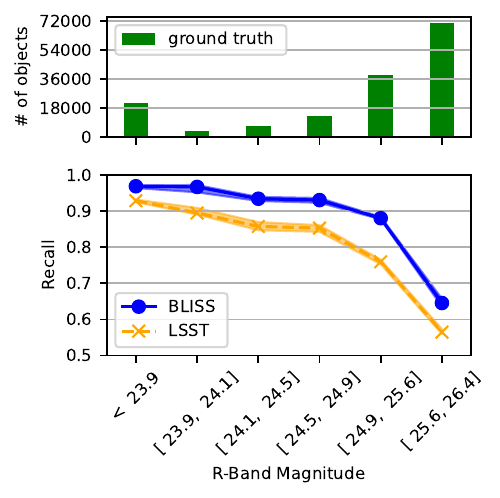}
    \caption{Detection precision (left) and recall (right). Each bar in the top row of bar plots represents the denominator for computing a point directly below it (in a line plot).}
    \label{fig:detection_performance}
\end{figure*}

\paragraph*{Precision and recall.}\quad
To generalize the concepts of precision and recall to spatial data, in keeping with \citet{portillo2017improved} and others, we define detection precision (purity) as the proportion of detections that are successfully matched with true objects and detection recall (completeness) as the proportion of objects that are successfully matched with detections.
\cref{fig:detection_performance} presents detection precision and recall as a function of magnitude for both BLISS and the LSST pipeline. The histogram in the top-right panel of \cref{fig:detection_performance} shows the total number of true objects in each r-band magnitude bin. Compared to LSST, BLISS successfully identifies a larger fraction of sources in each of the magnitude bins. Within the r-band magnitude range of 23.9 to 25.6, the recall of LSST decreases from 93\% to 58\%, and the recall of BLISS exhibits a similar drop from 97\% to 64\%.
The difference in recall between LSST and BLISS is largest in the second-to-last magnitude bin, where the difference in recall exceeds 10\%. Even for the faintest sources (magnitude $>$ 25.6), which are the most difficult to detect, BLISS outperforms the LSST pipeline in terms of recall by 8\%. 

The top-left panel in \cref{fig:detection_performance} shows the total number of detections from each method binned by magnitude.  The LSST pipeline yields more detections than BLISS in each bin, but BLISS exhibits systematically higher precision, particularly in the faintest magnitude bin.  
Further, BLISS differentiates faint objects and background noise well enough that it has many fewer spurious detections at the faintest magnitudes compared to the LSST pipeline. We can see examples of this in the top-left panel of \cref{fig:detection_example}: the four x-marks show positions where the LSST detection algorithm has confused noise with objects but BLISS has not.
We attribute the difference in detection performance for the faintest objects to limitations in the non-probabilistic approach to detection in the LSST pipeline, which leads to false detections and lower precision at the faintest magnitudes. Although the LSST pipeline detects more objects than BLISS throughout the magnitude bins considered, many of these detections are false detections, and thus the LSST pipeline nevertheless exhibits lower recall than BLISS.

\begin{figure*}[t]
    \includegraphics[width=0.47\textwidth]{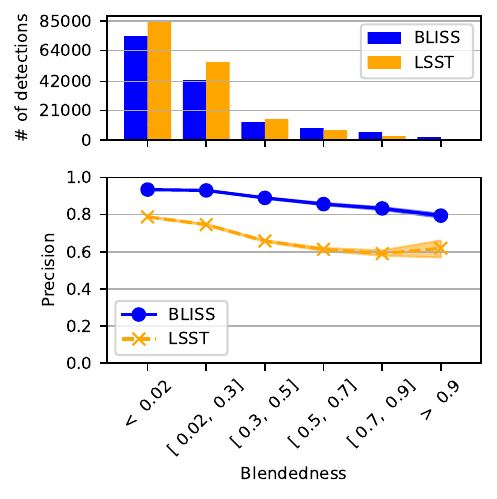}
    \hfill
    \includegraphics[width=0.47\textwidth]{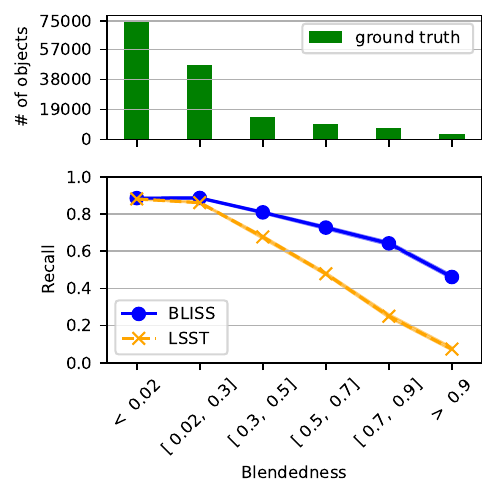}
    \caption{Detection precision (left) and recall (right) stratified by blendedness. Each bar in the bar plot represents the denominator for computing the point directly below it (in the line plot).}
    \label{fig:precision_recall_blendedness}
\end{figure*}

\paragraph*{Blended sources.}\quad
In addition to detecting fainter objects, BLISS demonstrates the ability to detect multiple objects in relatively crowded regions where blending occurs. The left plot of \cref{fig:precision_recall_blendedness} shows detection precision of BLISS and the LSST pipeline stratified by blendedness (as defined in \cref{blending_units}). BLISS gets better detection precision for objects of any blendedness. Both methods have similar recall (shown in the right plot) for the least blended objects.  However, BLISS's recall is markedly better for objects with blendedness in excess of $0.5$.
Examples of this appear in the top-right, bottom-left, and bottom-right panels of \cref{fig:detection_example}, which show light sources within 4 pixels of each other. In these three panels, BLISS successfully detects the centroids of multiple objects in each blend, identifying the 3, 2, and 3 objects, respectively. For the most severe blends, the LSST pipeline appears to mistake regions between objects for objects, as is illustrated by the x-marks between the circles.

\paragraph*{Centroid posterior calibration.}\quad
The first two rows of \cref{tab:credible_interval_coverage} report the calibration of our posterior approximation of objects' centroid positions.
For all coverage levels, the credible intervals were somewhat wider than the target.
For example, the credible interval coverage for a target of 90\% in the horizontal and vertical positions is $94\%$, indicating moderate overdispersion.
We discuss this overdispersion in \cref{sec:discuss-posteriors}.

\begin{figure*}
    \centering
    \includegraphics[width=0.43\textwidth]{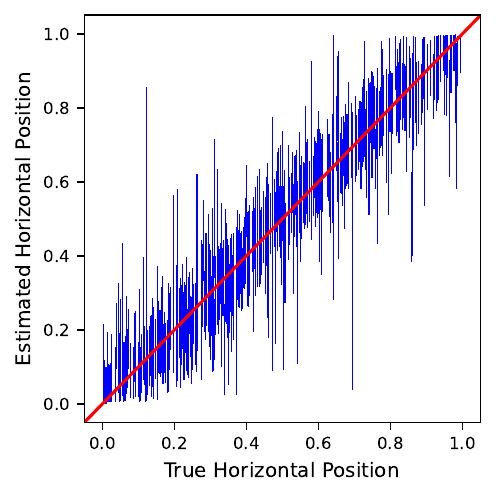}
    \hspace{3em}
    \includegraphics[width=0.43\textwidth]{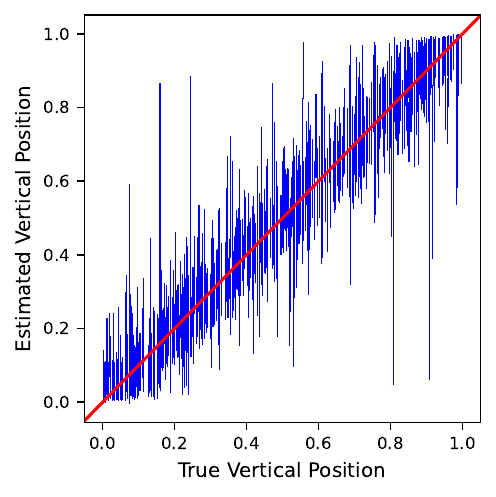}
    \caption{Highest-density 90\% credible intervals for the light source positions, formed according to the variational distribution and plotted against ground truth. Each interval represents a particular light source in a random sample of 470 light sources from our DC2 test set. The 45-degree line, which indicates perfect estimation, passes through the 94.0\% and 93.8\% of intervals for horizontal and vertical positions, respectively.
    }
    \label{fig:position_calibration}
\end{figure*}

\cref{fig:position_calibration} shows the credible intervals for each source without aggregation, providing a complementary evaluation of posterior calibration. It illustrates that credible intervals are fairly tight around the ground truth (that is, the 45-degree line) and that credible interval coverage is high across the range of positions.

\begin{table*}
    \centering
    \begin{tabular}{ccccc}
    \toprule
    \textbf{Name} & \textbf{90\% Coverage} & \textbf{80\% Coverage} & \textbf{70\% Coverage} & \textbf{50\% Coverage} \\ \midrule
    horizontal position & 0.944 & 0.882 & 0.811 & 0.644 \\
    vertical position  & 0.937 & 0.879 & 0.810 & 0.639  \\
\midrule
    star flux $u$ & 0.924 & 0.822 & 0.729 & 0.530 \\
    star flux $g$ & 0.939 & 0.838 & 0.755 & 0.557 \\
    star flux $r$ & 0.921 & 0.832 & 0.746 & 0.545 \\
    star flux $i$ & 0.928 & 0.830 & 0.754 & 0.569 \\
    star flux $z$ & 0.934 & 0.826 & 0.734 & 0.547 \\ 
    star flux $y$ & 0.922 & 0.818 & 0.717 & 0.520 \\
    galaxy flux $u$ & 0.912 & 0.813 & 0.718 & 0.512 \\
    galaxy flux $g$ & 0.913 & 0.820 & 0.733 & 0.533 \\
    galaxy flux $r$ & 0.918 & 0.818 & 0.732 & 0.523\\
    galaxy flux $i$ & 0.914 & 0.815 & 0.721 & 0.534\\
    galaxy flux $z$ & 0.922 & 0.812 & 0.716 & 0.522\\
    galaxy flux $y$ & 0.901 & 0.814 & 0.713 & 0.503 \\
\midrule
    ellipticity $g_1$ & 0.740 & 0.690 & 0.574 & 0.389  \\
    ellipticity $g_2$ & 0.770 & 0.693 & 0.590 & 0.412 \\
    \bottomrule
    \end{tabular}%
    \vspace{3pt}
    \caption{The frequentist coverage of highest-density credible intervals, formed based on the variational distribution. We consider four levels of coverage.}
    \label{tab:credible_interval_coverage}
\end{table*}

\subsection{Flux estimation}
Flux estimation is scientifically important because it allows astronomers to quantify the energy received from celestial objects, enabling the determination of their intrinsic properties such as luminosity, temperature, and composition, and facilitating studies of phenomena like variable stars, exoplanet transits, and supernovae.

\begin{figure*}
\centering
    \centering
    \includegraphics[width=0.9\textwidth]{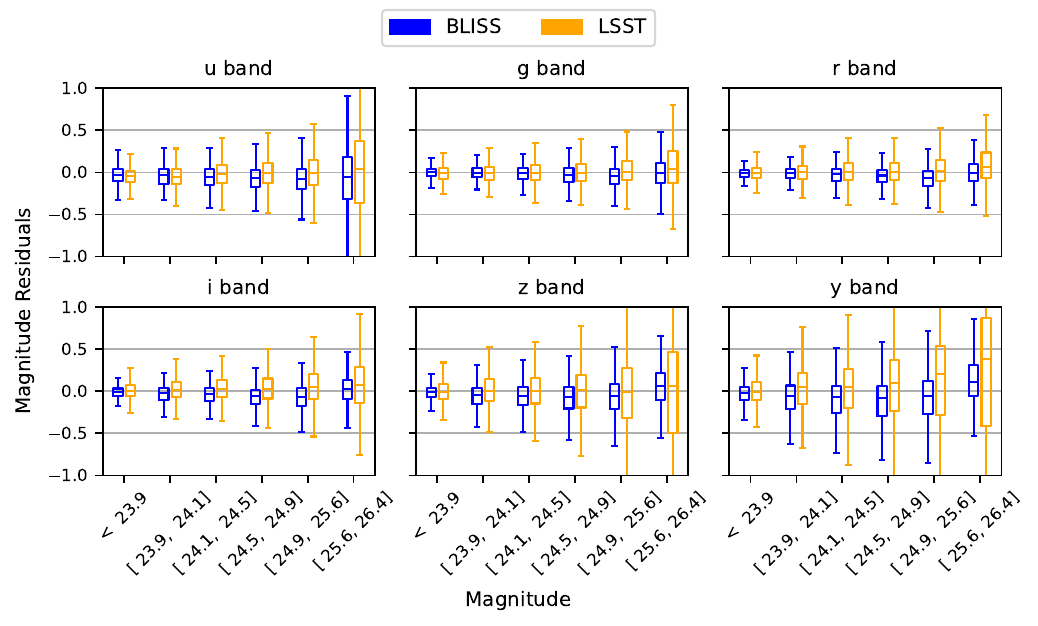}
    \caption{Magnitude residuals for the LSST pipeline and a point estimate based on the BLISS variational distribution. The magnitude residuals are defined as the difference between true magnitude and estimated magnitude. Each box plot shows the three quartiles, the maximum, and the minimum of the residual proportions.}
    \label{fig:mag_residuals}
\end{figure*}

\cref{fig:mag_residuals} shows the distribution of magnitude residuals in each of the six bands as a function of the true source magnitude for point estimates of magnitude from both BLISS and the LSST pipeline. We observe three trends in this comparison. First, the magnitude residuals (and thus the magnitude estimates) are less dispersed for brighter sources. This is likely due to the non-linear units of flux that we plot: the magnitude of brighter sources can be estimated with lower error because they are observed with less Poisson shot noise and interference from other sources as a proportion of the number of photoelectrons they contribute to the image. Second, the BLISS magnitude residuals have lower variance than those of the LSST pipeline for most magnitude bins and bands.
This difference is greatest for the faintest sources (magnitude $>$ 25.6) in the $y$ band, where the interquartile range for the LSST pipeline exceeds 1.0 magnitude, while the interquartile range for BLISS is less than 0.5 magnitude.
Third, both the BLISS and the LSST pipeline exhibit some bias in magnitude estimation, which appears to be roughly comparable in degree.

\begin{figure*}
\centering
    \centering
    \includegraphics[width=0.8\textwidth]{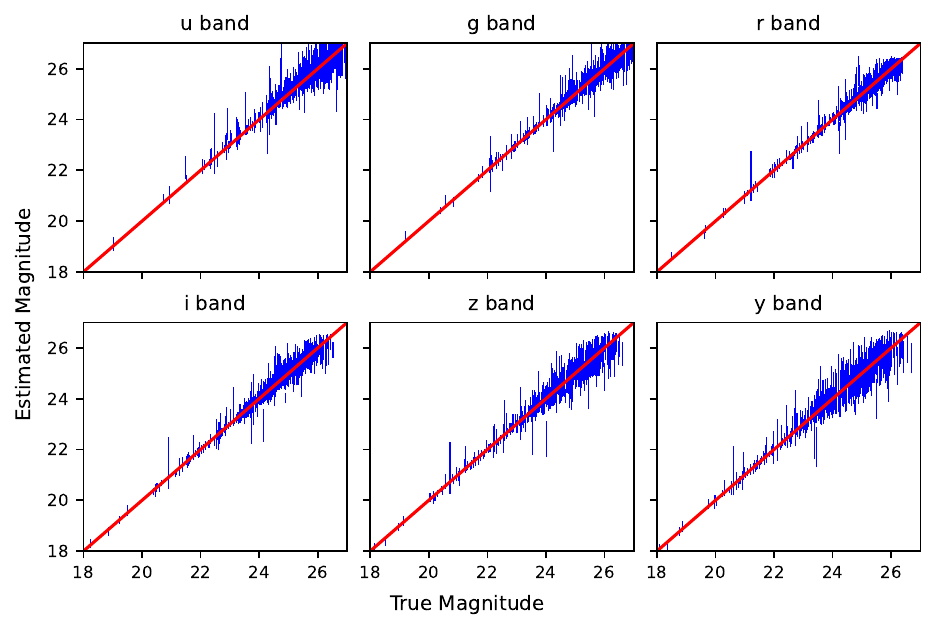}
    \caption{The highest-density 90\% credible intervals for the band-specific fluxes of light sources, constructed based on the variational distribution. Each interval corresponds to an individual light source from a random subset of 470 light sources in the test set. The 45-degree line represents perfect estimation accuracy.
    }
    \label{fig:flux_calibration}
\end{figure*}

\paragraph*{Posterior calibration.}\quad
The middle block of rows in \cref{tab:credible_interval_coverage} reports the frequentist coverage of credible intervals for flux.
For all coverage levels and all source types, the BLISS posterior approximation is somewhat overdispersed.
We discuss this overdispersion in \cref{sec:discuss-posteriors}.
\cref{fig:flux_calibration} shows the 90\% credible interval for sources from a subset of our test data, revealing that per-source credible intervals track the ground truth fairly closely for each filter band.

\subsection{Star/galaxy classification}
Classifying stars and galaxies in photometric surveys is essential for various scientific applications, including studies of galaxy formation and cosmology \citep{sevilla2018star}. In the field of weak lensing, identifying a pure sample of stars is crucial for fitting a model of the point spread function for use in inferring cosmic shear. Mistaking galaxies for stars could result in spurious correlations.

Star/galaxy separation is challenging because many light sources are faint and blended.
Further, once convolved with the PSF, some galaxies appear scarcely larger than stars. 
The LSST pipeline does not use machine learning to separate stars and galaxies. Instead, it compares the fit of an extended ``composite'' model to that of a point source \citep{bosch2018overview}

\begin{figure}
    \centering
    \includegraphics[width=0.45\textwidth]{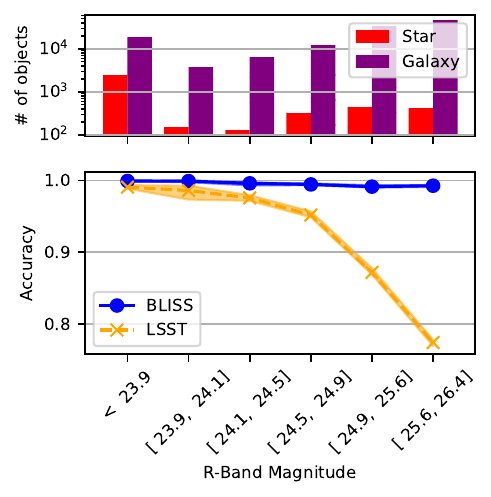}
    \caption{Star/galaxy classification accuracy for BLISS and LSST, stratified by true object magnitude. The bar plot (top) provides the counts of galaxies and stars in our test set, which are the denominators for calculating the accuracies shown in the line plot below.}
    \label{fig:classification_acc}
\end{figure}

In DC2, the star/galaxy classification problem is heavily imbalanced as the vast majority (97\%) of the objects are galaxies. The bar graph in the upper panel of \cref{fig:classification_acc} shows the counts of each object, stratified by magnitude.
The number of galaxies exceeds the number of stars in each magnitude bin by one or two orders of magnitude.

The overall star/galaxy classification accuracy for BLISS is 99.26\%.
The bottom panel of \cref{fig:classification_acc} shows the classification accuracy for each magnitude bin, for both BLISS and the LSST pipeline.
BLISS has systematically higher classification accuracy, successfully classifying at least 99.08\% of objects in each magnitude bin.
In contrast, for the faintest magnitude bin, the LSST galaxy classification accuracy is below 80\%.
The bootstrap confidence intervals indicate that this difference in performance is highly statistically significant.
Further, LSST galaxy classification accuracy systematically decreases as magnitude becomes fainter.

Appendix~D contains additional results for star/galaxy classification, including F1 scores and ROC curves for both bright and faint sources.

\subsection{Galaxy shape estimation}
Observed galaxy shapes provide a measurable signature of cosmic shear \citep{bacon2000detection}.
Measurement of many galaxies' shapes enables reconstructing mass maps that trace the large-scale dark matter distribution \citep{jeffrey2021dark}, constraining cosmological parameters such as $S_8$ \citep{asgari2021kids}, and inferring connections between galaxies and their dark matter halos \citep{clampitt2017galaxy}.

Galaxy ellipticities $g$ is a complex number that describes the elongation and orientation of a galaxy.
Its real part is \( g_1 = \frac{a - b}{a + b} \cos(2\varphi) \) and its imaginary part is \( g_2 = \frac{a - b}{a + b} \sin(2\varphi) \), where \( a \) and \( b \) are the lengths of the semi-major and semi-minor axes of the galaxy, respectively, and \( \varphi \) is the position angle of the major axis measured counterclockwise from a reference direction.

For the LSST pipeline, galaxy ellipticity is calculated as
\begin{align*}
    g = (I_{xx} - I_{yy} + 2i \times I_{xy}) / (I_{xx} + I_{yy} + 2 \times \sqrt{I_{xx} \times I_{yy} - I_{xy}^2}),
\end{align*}
where $I_{xx}$, $I_{yy}$, and $I_{xy}$ are the second moments in the ``truth match'' table (see Appendix~A).

We investigate the performance of BLISS and the LSST pipeline's ellipticity estimation stratified by galaxy magnitude (bright vs. faint) and blendedness (blended vs. unblended).  Classification performance drops sharply around 23.9 magnitude, suggesting this may be an interesting threshold. Therefore, we illustrate the difference between different populations based on this cut.
We consider bright galaxies to be those with magnitudes less than 23.9 and faint galaxies to be those with magnitudes ranging from 23.9 to 24.5. 
We take unblended galaxies to be those with blendedness (as defined in \cref{blending_units}) less than 0.02 and blended galaxies to be those with blendedness between 0.02 and 0.3.

Galaxies with magnitudes greater than 24.5 or blendedness exceeding 0.3 are excluded from the analysis as they are not reliably detected by either BLISS or the LSST pipeline, suggesting they provide minimal ellipticity information.
In the DC2 dataset, 21.4\% of sources have r-band magnitudes brighter than 24.5, and 77.5\% have blendedness values below 0.3, with 20.4\% of sources satisfying both criteria simultaneously.

\begin{figure*}
\centering
    \centering
    \includegraphics[width=0.9\textwidth]{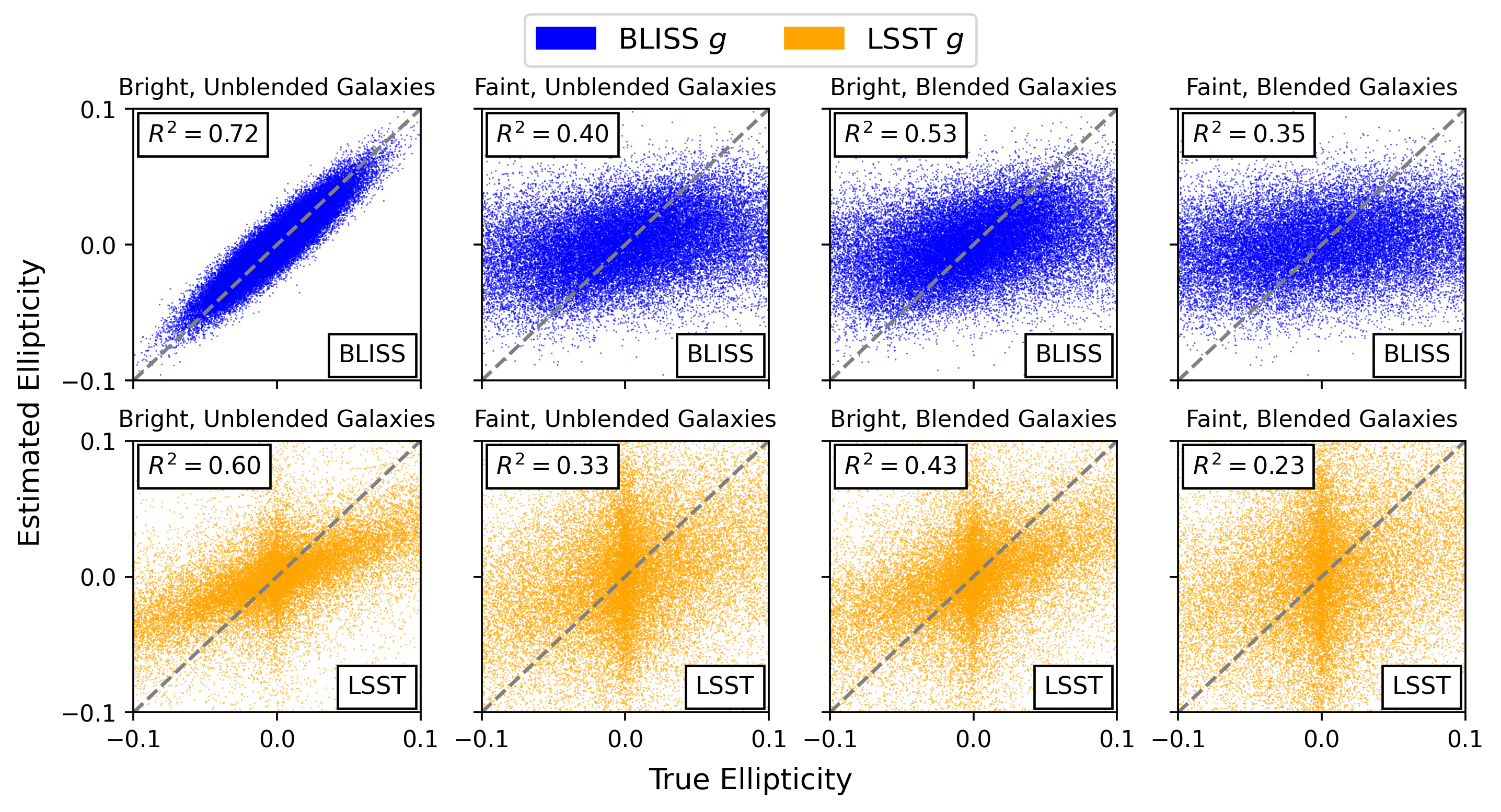}
    \caption{Galaxy ellipticity ($g_1$) estimates for BLISS and the LSST pipeline. Bright galaxies are defined as those with magnitudes less than 23.9, while faint galaxies possess magnitudes within the range of 23.9 to 24.5. Objects exhibiting a blendedness parameter below 0.02 are classified as unblended, whereas blended objects have blendedness between 0.02 and 0.3.}
    \label{fig:ellipticity_bliss_lsst}
\end{figure*}

\cref{fig:ellipticity_bliss_lsst} presents the distribution of estimated ellipticity versus true ellipticity, stratified by galaxy brightness and blendedness.
From \cref{fig:ellipticity_bliss_lsst}, we draw several conclusions. First, the variance and bias of ellipticity estimates increase as galaxy brightness decreases, as observed in the first two columns of the figure, where estimates for faint galaxies are biased toward zero and exhibit greater dispersion compared to those for bright and unblended galaxies. Second, higher blendedness in galaxies is associated with higher variance in ellipticity estimates, as demonstrated in the third column of the figure, where both BLISS and LSST produce estimates with greater errors for blended galaxies. Third, BLISS consistently outperforms LSST regardless of galaxy brightness and blendedness, with BLISS estimates exhibiting lower bias and variance across all four categories of galaxies depicted in the figure.

\paragraph*{Posterior calibration.}\quad
We first assess the posterior predictions of ellipticity parameters from BLISS using the frequentist coverage of credible intervals. 
As shown in \cref{tab:credible_interval_coverage}, the credible intervals for ellipticity are systematically narrower than the coverage they target, indicating overconfident posterior approximation.
For example, the credible intervals targeting 90\% coverage in fact cover only 74\% of the $g_1$ values and 77\% of the $g_2$ values.
We discuss this miscalibration further in \cref{sec:discuss-posteriors}.

\begin{figure*}
    \centering
    \includegraphics[width=0.47\textwidth, height=0.41\textwidth]{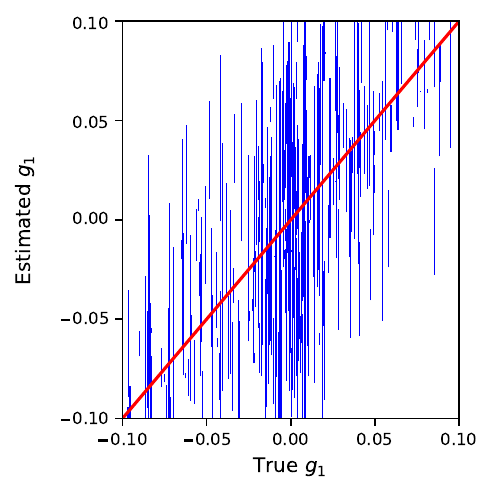}
    \hfill
    \includegraphics[width=0.47\textwidth, height=0.41\textwidth]{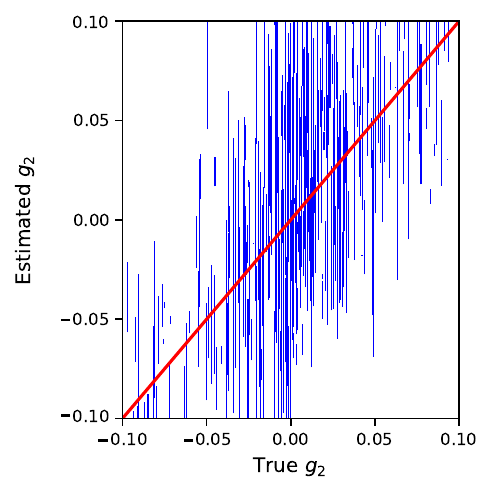}
    \caption{The highest-density 90\% credible intervals for ellipticity estimates, derived from the variational distribution. Each interval corresponds to a specific light source from a random sample of 470 light sources with magnitudes below 24.5 and blendedness less than 0.02 in our DC2 test set. The 45-degree line, representing perfect estimation, intersects with 82.3\% of the intervals for horizontal positions and 83.7\% for vertical positions.
    }
    \label{fig:ellipticity_calibration}
\end{figure*}

\cref{fig:ellipticity_calibration} reports credible intervals for individual galaxies. It indicates that ellipticity estimates track the ground truth.

\section{Discussion}
\label{sec:discussion}
When LSST begins full operations in 2026, it will usher in a new era of astronomical data collection, providing an unprecedented opportunity to advance our understanding of the cosmos. Science requirements dictate the precise measurement of billions of galaxies.
However, the high density of light sources in LSST images will exacerbate the limitations of existing algorithmic pipelines, which currently do not handle ambiguity in a statistically coherent way.

\subsection{The potential of NPE to improve accuracy}
\label{potential}
Deep learning has revolutionized image processing in many fields, consistently achieving much higher predictive accuracy than alternative methods. However, deep learning is conspicuously absent in traditional astronomical cataloging pipelines. Our results show that deep learning, used within an NPE framework, can dramatically increase the accuracy of astronomical catalogs. In relation to the planned LSST pipeline, our experiments demonstrate systematically better performance by NPE for the following tasks.
\begin{itemize}
    \item Light source detection: 58\% higher precision (0.79 vs 0.50) in combination with 10.3\% higher recall (0.64 vs 0.58) for faint light sources, as well as 686\% (0.48 vs 0.07) higher recall for heavily blended light sources.
    \item Flux (magnitude) measurement: a 56\% reduction in the absolute error for faint light sources in the lowest SNR band.
    \item Star/galaxy classification: 96\% fewer classification errors for faint light sources.
    \item Galaxy ellipticity estimation: a 60\% reduction in absolute error.
\end{itemize}
The LSST pipeline detects objects according to the peaks of flux in the images and relies on an algorithm with multi-band peak merging and thresholds for peak detection.  The LSST approach does not explicitly quantify uncertainty.  Our approach, based on a generative model, enables our algorithm to more easily distinguish faint sources from background noise. If the background has high levels of random noise, such as that in \cref{fig:detection_example}, it is ambiguous whether a small peak in flux corresponds to background noise or a very faint source.
In line with this, the most pronounced improvements are typically found in the faintest magnitude bin.  However, we found substantial improvements in essentially every magnitude and blendedness bin for each of the tasks we considered.

One caveat concerning these results is that no handful of metrics, including the ones we consider, can fully assess the relevance of the performance improvements to downstream scientific applications, which ultimately seek understanding of nature rather than quantifiable performance.
Nevertheless, these systematically lower error rates are encouraging.

\subsection{The potential of NPE to approximate the posteriors}
\label{sec:discuss-posteriors}
NPE imposes few restrictions on the form of the generative model and implicitly marginalizes over nuisance random variables. These capabilities allow us to perform inference under the correctly specified model of coadded multiband images (that is, the coaddition routine itself) and to consider multiple bands of images jointly.

Credible intervals based on our posterior approximations show our NPE approximations to be well calibrated, although slightly (several percent) too large.
Overdispersion is not necessarily a problem, as we typically do not expect perfect calibration from variational inference, which has restrictions in the form of the variational distribution that are not shared by the exact posterior.
In particular, our variational distribution is composed of simple parametric distributions. Because NPE minimizes the forward (a.k.a. exclusive) KL divergence, overdispersed approximations often result when the exact posterior cannot be recovered due to restrictions on the form of the approximation \citep[Ch. 10]{bishop2006pattern}.

If this overdispersion were problematic, our variational distribution could be made more flexible. For example, we could use mixture models \citep{bishop1994mixture} or normalizing flows \citep{papamakarios2021normalizing} to define our variational distribution.
The ease with which we can enhance the flexibility of our variational distribution used by NPE is not shared by traditional (ELBO-based) approaches to variational inference, which typically require the reparameterization trick or, for discrete variables, continuous relaxation. In contrast, NPE imposes almost no restrictions on the form of the variational distribution.

However, while we have a clear path to decreasing miscalibration through the introduction of successively more flexible variational distributions, this path likely will not converge to perfect calibration. 
Sensitivity analyses will be needed to assess whether the miscalibration is substantive for any particular task. Such analyses should take into account how catalogs are used for the task in question.
It often will not suffice to assess metrics that are averages across light sources, as we have done in this article, as many uses of catalogs instead involve population-level analysis \citep{schafer2015framework}.
In population-level analysis, slight miscalibration in source parameters (e.g., galaxy ellipticity) can compound, leading to large errors in the inference of population-level quantities.


\subsection{Potential model misspecification}
\label{misspec}

Our results highlight the performance achievable without model misspecification, as both our training and test data were generated by the DC2 simulator, which ensures that they follow the same distribution. Although the DC2 simulator offers a high level of realism, it remains a simulation. Hence, these results can be seen as the best-case scenario for the performance of neural posterior estimation (NPE). However, when NPE is applied to real LSST data, the inference network will have been trained with data sampled from a model that is at least somewhat misspecified.

The DC2 simulator, although realistic, may not fully capture the diversity needed to reflect our gaps in understanding the Universe. It was designed for algorithm validation rather than training for application to real data. For instance, it employs the Outer Rim simulation, which uses a single parameterization of the $\Lambda$CDM model of cosmology, a widely accepted framework that describes the Universe as composed of dark energy (denoted by $\Lambda$), cold dark matter (CDM), and ordinary matter.  Since determining the true parameters of this cosmological model is an open area of research, any particular parameter values selected by practitioners are almost certainly incorrect to some extent.

NPE can be particularly vulnerable to model misspecification due to its reliance on simulated data. If the simulator is misspecified due to uncertainty about the true data generating process, the resulting inference network might be applied to out-of-distribution settings, necessitating extrapolation by the neural network. In this setting, typical neural networks are known to perform poorly \citep{shen2024engression}.

Fortunately, several domain adaptation techniques can mitigate these issues. One method, proposed by \cite{ward2022robust}, introduces a simple parametric error model for the observed data, which is then stochastically inverted. Another method, by \cite{gloeckler2023adversarial}, involves regularizing the Fisher information matrix of the model parameter with respect to the data, inspired by considerations of adversarial attacks. Further, \cite{huang2023learning} suggested adding a regularizer to the NPE objective based on the maximum mean difference (MMD). A further technique by \cite{wehenkel2024addressing} employs a small calibration set consisting of pairs of real-world observations and ground truth, which for our work could potentially come from more precise but less available instruments that produce smaller volumes of data (e.g., spectrographs).

These domain adaptation techniques seem useful for alleviating the susceptibility of NPE to model misspecification as long as the amount of regularization they provide is mild. But they implicitly change the model under which we are performing inference in ways that are not transparent. 
Therefore, they do not seem like a reliable basis for overcoming the fundamental issue of model misspecification.

\subsection{A path forward: nonparametric modeling}
\label{nonparametric}
Best practice for a Bayesian workflow involves iterative development guided by predictive checks of models with varying complexities \citep{gelman2020bayesian}. This iterative process allows analysts to progressively refine their models, ultimately addressing various sources of model misspecification. By incrementally adding uncertainty to a simulator based on predictive checks and domain expertise, we can reduce the degree of model misspecification.

At least three components of the simulator may require generalization. The galaxy model may need to account for irregular structures that extend beyond simple bulge-and-disk formations.
The point spread function (PSF) may require adjustments for color dependence. The background model may need to incorporate structured variations. Further, there are possible model misspecifications, such as the assumed SED of simulated light sources.

Although this article focuses on DC2, which was simulated using GalSim (with imSim extensions), GalSim is primarily designed to simulate images used for validation rather than training models. 

There has been work developing deep generative models of galaxy images, and indeed, the models may be less misspecified \citep{regier2015deep, lanusse2021deep, smith2022realistic}. However, it is unclear how to train these models as training requires access to segmented galaxy images, which cannot be reliably found without an accurate galaxy model. 
In contrast, AstroPHOT \citep{stone2023astrophot} offers a more flexible model, specifically designed to fit images. AstroPHOT provides nonparametric component models, including Fourier- and spline-based object models, composite models, and extensible modules. This flexibility may make AstroPHOT a more appropriate simulator than GalSim for the future development of our inference method.

As the model becomes more flexible and parameter-rich, however, specifying a prior becomes more challenging due to the decreased interpretability of the parameters and the more complex dependency structure among them. Additionally, adding flexibility to the model can lead to the possibility of non-identifiable or weakly identifiable parameters. Since we lack a priori knowledge about the dependence structure, the prior needs to be diffuse and hierarchical with global random variables to adapt based on the observed data. This poses a significant challenge because we do not know what form of hierarchical prior to use.

The expectation-maximization (EM) algorithm could provide an alternative to explicitly setting a hierarchical prior. 
Although EM is not globally convergent, it improves the marginal likelihood of the data with each iteration.
Further, for identifiable model parameters (or identifiability transformations of model parameters), with strong regularity conditions, the EM algorithm can be a consistent estimation procedure \citep{balakrishnan2017statistical}.

NPE can approximate the E-step in an EM algorithm.
In this approach, NPE infers posteriors for the local random variables, while the M-step updates the model parameters based on the current posterior approximation. Whether this procedure inherits the desirable properties of EM depends on the accuracy of the approximation learned by NPE.
The global convergence properties of NPE \citep{mcnamara2024globally} and the flexibility of modern variational approximations, such as normalizing flows and conditional diffusion models \citep{sharrocksequential}, enable NPE to solve the E-step almost exactly, which is important for retaining the theoretical properties of the EM algorithm.
The proposed work can hence be viewed as a powerful approximate solution to the E-step of the EM algorithm, which, with conditions, can be a consistent estimation procedure.

The traditional cataloging workflow in astronomy has no similar path to providing consistent estimation, nor has it even been considered in these terms; hard thresholds and a lack of uncertainty quantification make it difficult to see how a path could exist.
In contrast, the proposed method could serve as a subroutine in an approximately consistent estimation procedure, to more faithfully characterize the Universe.

\section*{Author Contributions}
YD conducted the numerical experiments and created the figures.
YD, XL, and JR extended the BLISS software to catalog DC2.
CA and JR conceived of the project and supervised the work.
YD, XL, CA, and JR wrote the text of this article.

\section*{Acknowledgements}
This material is based on work supported by the National Science Foundation under Grant No. 2209720 and the U.S. Department of Energy, Office of Science, Office of High Energy Physics under Award Number DE-SC0023714.

This paper has undergone internal review in the LSST Dark Energy Science Collaboration. The internal reviewers were Chad Schafer and Grant Merz. 
The DESC acknowledges ongoing support from the Institut National de 
Physique Nucl\'eaire et de Physique des Particules in France; the 
Science \& Technology Facilities Council in the United Kingdom; and the
Department of Energy and the LSST Discovery Alliance
in the United States.  DESC uses resources of the IN2P3 
Computing Center (CC-IN2P3--Lyon/Villeurbanne - France) funded by the 
Centre National de la Recherche Scientifique; the National Energy 
Research Scientific Computing Center, a DOE Office of Science User 
Facility supported by the Office of Science of the U.S.\ Department of
Energy under Contract No.\ DE-AC02-05CH11231; STFC DiRAC HPC Facilities, 
funded by UK BEIS National E-infrastructure capital grants; and the UK 
particle physics grid, supported by the GridPP Collaboration.  This 
work was performed in part under DOE Contract DE-AC02-76SF00515.

\section*{Software Availability}

Our software is available in a public GitHub repository: \url{https://github.com/prob-ml/bliss}.
The specific code for reproducing the plots and tables in this article can be found in the \texttt{case\_studies/dc2\_cataloging} directory.



\bibliography{reference}{}
\bibliographystyle{aasjournal}

\appendix

\section{Interpreting DC2 catalogs}
\label{appendixa}
\citet{abolfathi2021desc} provides a comprehensive guide to the DC2 dataset. Here, we identify and describe just the subset of DC2 fields used in our comparisons.
\begin{table}[ht!]
\centering
\resizebox{\textwidth}{!}{%
\begin{tabular}{lllll}
\toprule
 & \textbf{Variable Name} & \textbf{Table} & \textbf{Units} & \textbf{Description} \\

\midrule
\multirow{2}{*}{\textbf{Direction}} & \texttt{ra} & \multirow{2}{*}{Truth} & degree & right ascension (lensed) \\
 & \texttt{dec} &  & degree & declination (lensed) \\

\midrule
\textbf{Source Type} & \texttt{truth\_type} & Truth & - & whether object is a star or galaxy \\

\midrule
\textbf{Photometry} & \texttt{flux\_<band>} & Truth & nJy & static flux value in a band \\

\midrule
\multirow{8}{*}{\textbf{Galaxy Shape}} & \texttt{ellipticity\_1\_true\_dc2} & \multirow{8}{*}{CosmoDC2} & - & ellipticity component 1, for galaxy, not lensed \\
 & \texttt{ellipticity\_2\_true\_dc2} &  & - & ellipticity component 2, for galaxy, not lensed \\
 & \texttt{size\_disk\_true} &  & arcsec & disk half-light radius (of major axis), not lensed \\
 & \texttt{size\_bulge\_true} &  & arcsec & bulge half-light radius (of major axis), not lensed \\
 & \texttt{size\_minor\_disk\_true} &  & arcsec & disk half-light radius (of minor axis), not lensed \\
 & \texttt{size\_minor\_bulge\_true} &  & arcsec & bulge half-light radius (of minor axis), not lensed \\
 & \texttt{position\_angle\_true} &  & degree & position angle ($\arctan(E_2/E_1)$), for galaxy, not lensed \\
 & \texttt{bulge\_to\_total\_ratio\_<band>} &  & - & bulge-to-total luminosity ratio in a band \\
\bottomrule
\end{tabular}%
}
\caption{DC2 ground-truth parameters. The ``Table'' column refers to the table from which the variable comes, with ``Truth'' corresponding to \texttt{desc\_dc2\_run2.2i\_dr6\_truth} and ``CosmoDC2'' corresponding to \texttt{desc\_cosmodc2}.\vspace{-15pt}
}
\label{tab:exp_var}
\end{table}

\begin{table}[ht!]
\centering
\resizebox{\textwidth}{!}{%
\begin{tabular}{lllll}
\toprule
 & \textbf{Variable Name} & \textbf{Table} & \textbf{Units} & \textbf{Description}\\

\midrule
\multirow{2}{*}{\textbf{Direction}} & \texttt{ra} & \multirow{2}{*}{Truth Match} & degree & right ascension (lensed) \\
 & \texttt{dec} &  & degree & declination (lensed)  \\

\midrule
\textbf{Source Type} & \texttt{extendedness} & Truth Match & - & whether object is a star or galaxy  \\

\midrule
\multirow{2}{*}{\textbf{Photometry}} & \texttt{cModelFlux\_<band>} & Truth Match & nJy & galaxy flux in the specified band \\
& \texttt{psFlux\_<band>} & Truth Match & nJy & star flux in the specified band \\

\midrule
\multirow{3}{*}{\textbf{Galaxy Shape}} & \texttt{Ixx\_pixel} & \multirow{3}{*}{Truth Match} & - & second moment (xx) of source intensity \\
& \texttt{Iyy\_pixel } &  & - & second moment (yy) of source intensity \\
& \texttt{Ixy\_pixel} &  & - & second moment (xy) of source intensity \\

\midrule
\multirow{4}{*}{\textbf{PSF}} & \texttt{IxxPSF\_pixel\_<band>} & \multirow{4}{*}{Truth Match} & pixel$^2$ & second moment (xx) of the PSF in a band \\
 & \texttt{IxyPSF\_pixel\_<band>} & & pixel$^2$ & second moment (xy) of the PSF in a band  \\
 & \texttt{IyyPSD\_pixel\_<band>} & & pixel$^2$ & second moment (yy) of the PSF in a band  \\
 & \texttt{psf\_fwhm\_<band>} &  & arcsec & PSF full width at half maximum  \\

\midrule
\textbf{Blendedness} & \texttt{blendedness} & Truth Match & - & measure of blendedness in [0, 1]; 0 = unblended  \\
\bottomrule
\end{tabular}%
}
\caption{Parameter estimates from the LSST pipeline. ``Truth Match'' refers to the \texttt{desc\_dc2\_run2.2i\_dr6\_object\_with\_truth\_match} table.
}
\label{tab:lsstestimated}
\end{table}

\clearpage
\section{An Algorithmic Perspective on the BLISS Pipeline}\label{sec:app_bliss_pipeline_details}

In the main text, our training and inference procedures are presented as following from our choice to use neural posterior estimation (NPE) with a particular variational family, and from ``turning the Bayesian crank.'' Here, we give a complementary perspective that is more algorithmic and more focused on data structures.

Our proposed method, BLISS, takes multi-channel images as input and produces catalogs as output.
Internally, BLISS represents catalogs in a ``tiled'' format (Section~\ref{sec:tile-catalogs}).
To infer a complete catalog for an image, owing to its autoregressive structure, the BLISS network (Section~\ref{sec:recurrent-cnn}) requires multiple forward passes by some of its modules, with each making inferences for a different set of tiles in the context of varying conditioning information, which is also provided as a (partially masked) tile catalog.
During training, the BLISS network is challenged with inferring (probabilistically infilling) randomly selected tiles, given the image and the truth about the objects in a disjoint random set of tiles; in contrast, during inference, sampling the variational distribution follows a deterministic pattern (Section~\ref{sec:masking-patterns}).
Training is computationally intensive, requiring roughly one day on a single GPU, whereas inference is roughly as fast as data loading, which implies that the proposed approach can readily be scaled to process the entire LSST coadded image set (Section~\ref{sec:computational-reqs}).

\subsection{Tile Catalogs}
\label{sec:tile-catalogs}
A tile catalog is a particular format of an astronomical catalog, which we designed. Whereas a traditional astronomical catalog has two dimensions, with one row per light source and one column per property, a tile catalog has four dimensions. The first two index small spatial regions (tiles), the next indexes sources whose centroid is within the tile, and the final dimension indexes the properties of each source. Tile catalogs are padded to have a fixed size, and paired with another tensor that indicates which entries are padded.
Thus, despite representing a variable number of sources per tile, they can be represented as dense tensors rather than as ragged arrays.

\begin{figure}
    \centering
    \includegraphics[width=\textwidth, trim=35 350 35 5, clip]{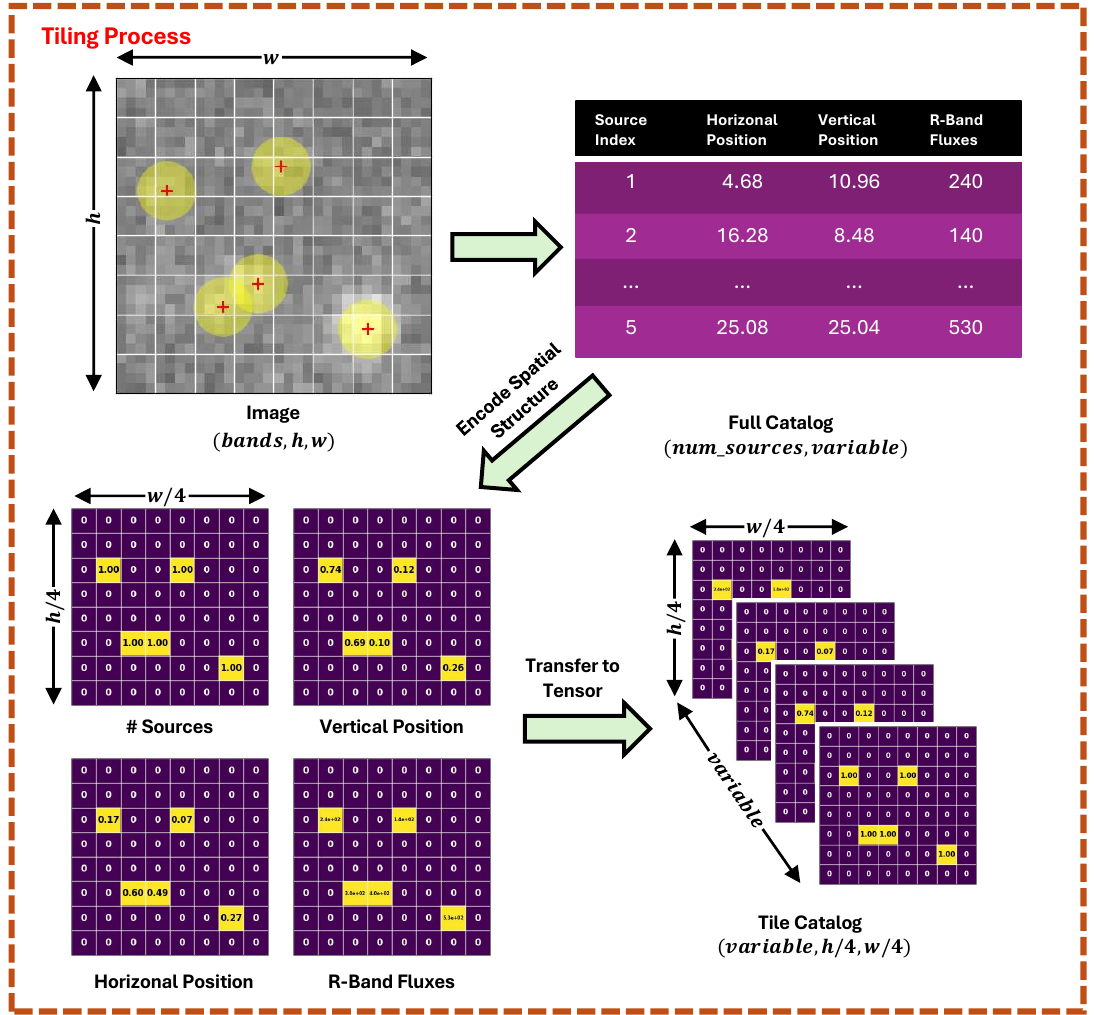}
    \caption{The tiling process. For illustration purposes, we show the conversion of a 32$\times$32 image and highlight four selected variables. Source locations are marked with `+' symbols. Note that while the full catalog stores absolute pixel coordinates, the tile catalog stores relative positions within each tile.}
    \label{fig:tiling_process}
\end{figure}

Tile catalogs, in contrast to traditional catalogs, can be directly outputted by fully convolutional neural networks because the lengths of their first two dimensions are proportionate to the height and width of the input image.
With a fully convolutional network, the receptive field for each tile in the output is centered at the region described by the tile. Whereas the tile catalog is more convenient as a neural network response (during training or inference), the traditional catalog format is more useful for reporting and for data augmentation. Fortunately, using sparse vector operations, it is possible to efficiently convert a tile catalog to a traditional catalog, or vice versa, without moving the catalog data off the GPU. 

Figure~\ref{fig:tiling_process} illustrates the relationships among a 32$\times$32 pixel image and its catalog in both tile- and traditional-catalog formats. In the example, there are five sources, each represented as a row in the full catalog, with attributes such as position and flux stored in corresponding columns. To generate the tile catalog, each source is assigned to the 4$\times$4-pixel tile that contains its center. For instance, a source located at pixel coordinates (10.96, 4.68) lies within tile (2, 1). Consequently, we increment the count in the ``\# Sources'' array at position (2, 1), and populate the other attribute arrays at the same tile coordinates with the respective values for this source. Before storing the pixel positions, we convert them to relative coordinates within the tile. Specifically, the relative position of (10.96, 4.68) in tile (2, 1) becomes (0.74, 0.17), where $(10.96 \bmod 4) / 4 = 0.74$ and $(4.68 \bmod 4) / 4 = 0.17$. This process is repeated for all sources, and the resulting two-dimensional arrays are stacked to form the final tile catalog.

\subsection{Recurrent CNN architecture}
\label{sec:recurrent-cnn}
\begin{figure}
    \centering
    \includegraphics[width=\textwidth, height=0.82\textheight, trim=35 250 35 5, clip]{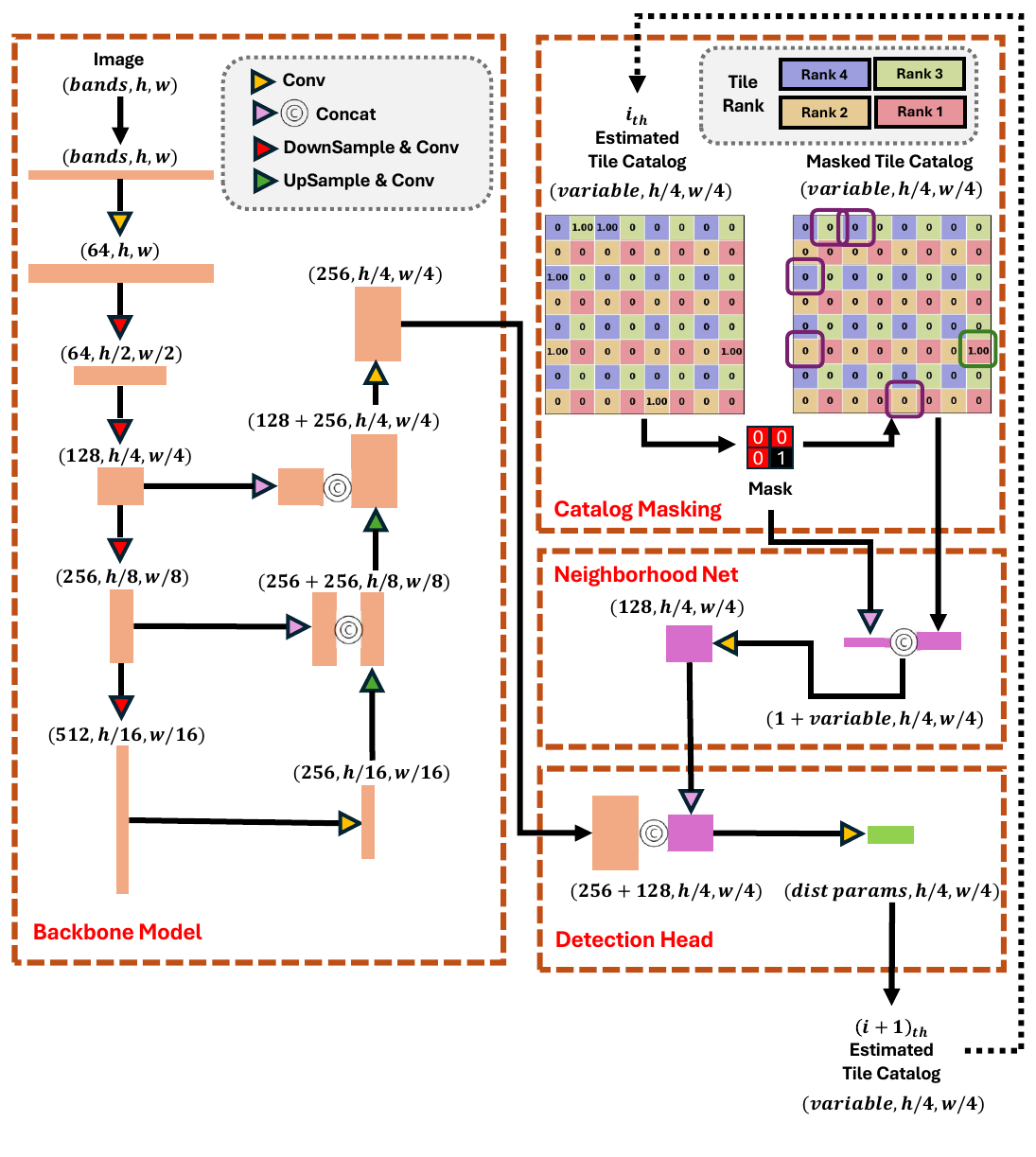}
    \caption{The BLISS inference workflow. Without loss of generality, we illustrate the second inference step ($i=2$). Neural network operations are indicated by arrows of different colors. Tensor shapes follow the (c, h, w) format, denoting channels, height, and width. In the catalog masking box, we use the ``\# Sources'' array from the tile catalog for illustration. Tiles of different ranks are shown using distinct colors.}
    \label{fig:inference_plot}
\end{figure}

During inference, BLISS approximates the posterior distribution of source attributes for each tile. Figure~\ref{fig:inference_plot} depicts this inference workflow. The model takes an astronomical image as input and outputs the parameters of posterior approximations. A U-shaped backbone network, adapted from \citet{ultralytics2021yolov5}, extracts tile-level image features. Simultaneously, the neighborhood network models spatial dependencies via an autoregressive process by conditioning on a masked version of the previously inferred tile catalog. At inference step $i$, only the tiles with ranks lower than $i$ are revealed to the neighborhood network. The catalog masking box of Figure~\ref{fig:inference_plot} illustrates this process for inferring the catalogs of the rank-2 tiles. Here, only rank-1 tiles (highlighted in red) are unmasked. The masked tiles are outlined in purple rectangles, and the preserved ones in green rectangles. The extracted image features are concatenated with the neighborhood network's output and passed to the detection head, which predicts 22 distribution parameters per tile (described in the main text) via a 1$\times$1 convolutional layer. If the current step is not the final one, samples drawn from the predicted posterior distributions are used to form a new tile catalog, which serves as input for the neighborhood network in the next step.

\subsection{Masking patterns}
\label{sec:masking-patterns}
\begin{figure}
    \centering
    \includegraphics[width=\textwidth, trim=35 520 35 5, clip]{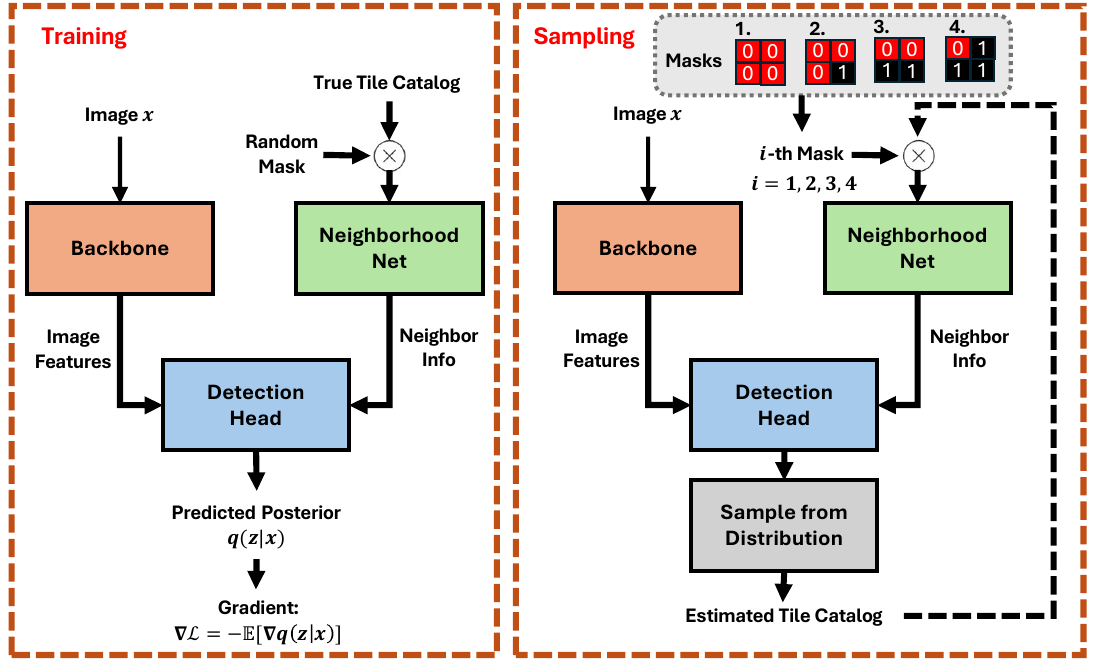}
    \caption{Training and sampling in BLISS. ``Sample from Distribution'' indicates the process of drawing samples from the estimated posterior distributions.}
    \label{fig:train_sample_plot}
\end{figure}

Figure~\ref{fig:train_sample_plot} provides an overview of both training and inference in BLISS. During training, the neighborhood network is conditioned on a partially observed ground truth tile catalog. Each training example is masked using one of 15 possible patterns. At inference time, the first step uses an all-zero mask, effectively providing an empty tile catalog as context; predictions are made independently of neighboring tiles. In the second step, the previously estimated tile catalog is inputted, with only rank-1 tiles unmasked. This allows the model to refine its predictions for rank-2, rank-3, and rank-4 tiles based on information from rank-1. The third and fourth steps proceed similarly, incrementally incorporating spatial dependencies through selective unmasking.

The stochasticity in these masking patterns is in addition to more traditional data augmentation. We also apply random 90-degree rotations, flips, and small horizontal and vertical shifts to both the images and the catalogs. This helps us to avoid overfitting despite the modest size (by neural network standards) of the DC2 coadd image set.

\subsection{Computational requirements}
\label{sec:computational-reqs}
The DC2 coadd image set covers 2 tracts (2.56 square degrees each), each consisting of 49 patches arranged in a 7$\times$7 grid. For each of these 98 patches, there is a coadded image with 6 channels and 4100$\times$4100 pixels per channel.
On a modern GPU, given the depth of our network, these images are too large to be used for training, even with a batch size of one: typical GPUs do not have enough memory to store one of these images, the intermediate results of performing a forward pass with it, and the gradients produced during the backward pass.
Therefore, we split these high-resolution images into 6$\times$80$\times$80-pixel sub-images, and train with minibatches of 32 sub-images.

Training the inference network is computationally intensive (although it is an upfront cost that should not change greatly with the size of the dataset). With one NVIDIA RTX 2080 Ti GPU, a batch size of 32, and a learning rate of 0.001, the optimizer converges in approximately 22 hours over 30 epochs. Each epoch requires around 45 minutes to complete, with GPU memory utilization reaching 7074 MB.

In contrast, inference is almost instant, taking roughly as long as the data loading itself: the network performs posterior inference for 100 images, each $80 \times 80$ pixels, per second. This runtime implies that we can process a patch in 26 seconds, and an entire tract in less than 22 minutes.
The runtime bottleneck is data loading rather than neural network computation. To minimize this limitation, we load data in background threads while the GPU simultaneously processes already-loaded images.
As backward propagation is not required during inference, GPU memory utilization is only 706 MB.

To apply BLISS to future LSST data releases, the inference runtime is more relevant to training time.
Once the network is trained, the total runtime on actual LSST data would simply scale with the number of image patches.
Given the inference runtime reported above and a projected LSST survey footprint of 18,000 square degrees, we can further project that processing one coadd for each LSST tract would require 107 days with a single GPU.
This workload is easily parallelized across nodes.

In processing all LSST coadds, the BLISS output catalog itself would be quite large on disk, and perhaps multiple times larger than the LSST pipeline's catalog if we stored multiple posterior samples per image.
For analyses of large populations of astronomical objects, a catalog based on a single posterior sample may be adequate, but for analyses of individual or few sources, more samples would typically be desired.
For these analyses of a small number of sources, rerunning BLISS inference (with a pretrained model) may be more appealing than using precomputed catalogs, which scale in size with the number of samples.

\clearpage
\section{Different Matching Threshold}
\label{apdx:diff_matching}

\begin{figure}[H]
    \centering
    \includegraphics[width=\textwidth]{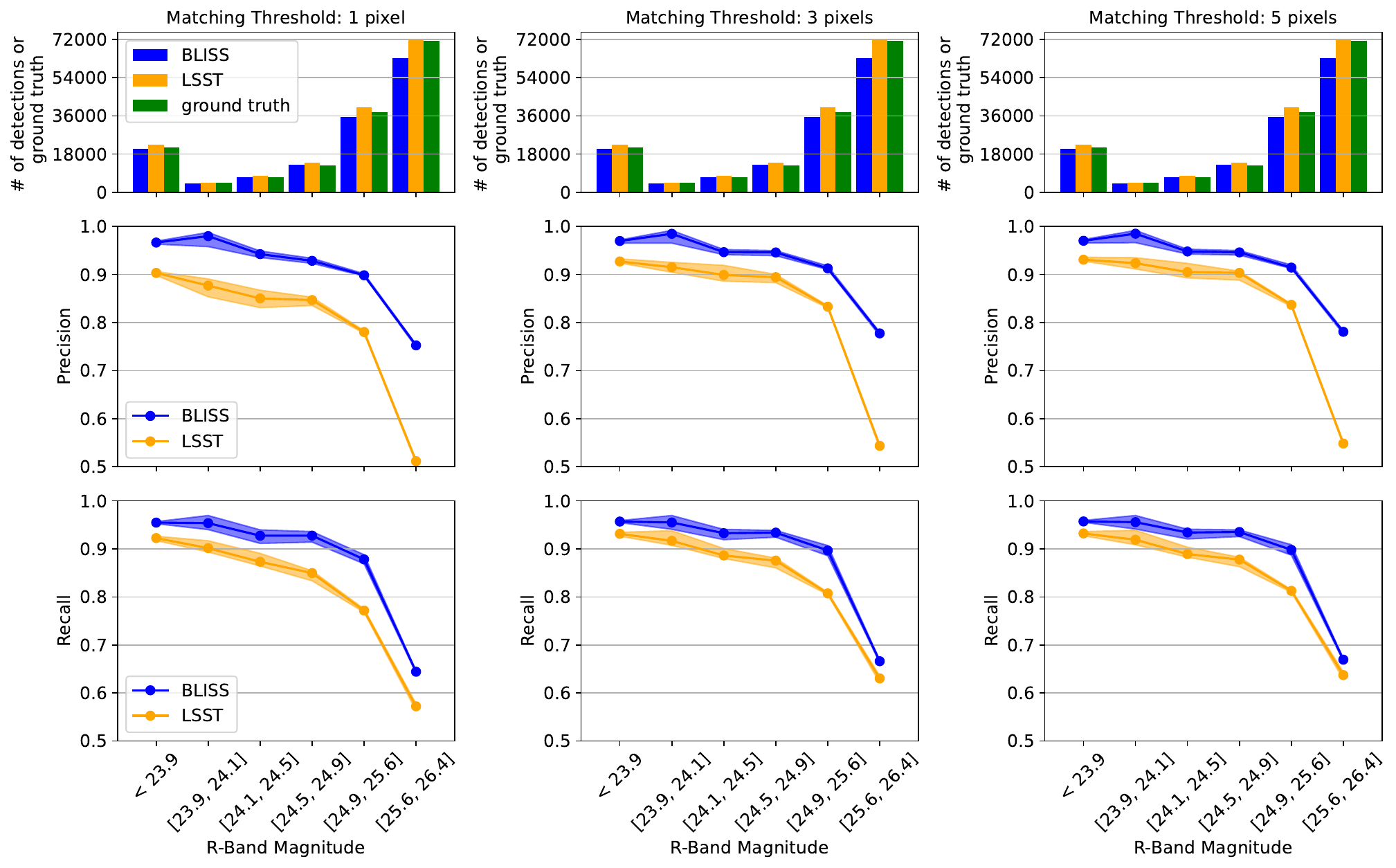}
    \caption{Detection precision and recall for various matching thresholds. Each column reports the results for a particular threshold: 1 pixel (left), 3 pixels (center), and 5 pixels (right).}
    \label{fig:diff_match}
\end{figure}

In the main text, a detected object could only match a true object if their centroids were within one pixel of each other.
Figure~\ref{fig:diff_match} shows that while looser matching thresholds improve both detection precision and recall for both BLISS and LSST, BLISS outperforms the LSST pipeline regardless of which threshold is used. BLISS maintains a higher recall than the LSST pipeline at a 3-pixel matching threshold, and our findings for precision are similar. The right column of subplots shows that thresholds beyond 3 pixels yield only marginal gains in recall and precision. This suggests that BLISS outperforms the LSST pipeline in detection accuracy irrespective of the matching threshold.

\clearpage
\section{Additional results for star/galaxy classification}
\label{sec:more-classification}

In the main text, we report the accuracy of the star/galaxy classification. F1 scores and ROC curves provided complementary assessments of classification performance.

\paragraph*{F1 score.}\quad
The F1 score, which is the harmonic mean of precision and recall, provides a combined score of classification performance and is meaningful even in the presence of class imbalance. To compute precision and recall, and hence to compute the F1 score, one class needs to be designated as the ``positive'' class. We computed the F1 score both with stars considered the positive class and with galaxies considered the positive class.

\begin{figure}[h]
    \centering
    \includegraphics[width=0.47\textwidth]{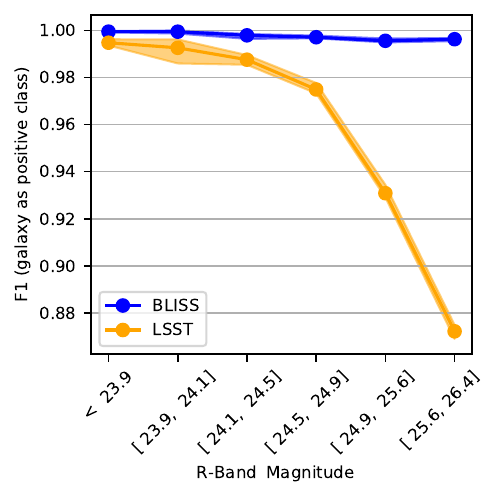}
    \hfill
    \includegraphics[width=0.47\textwidth]{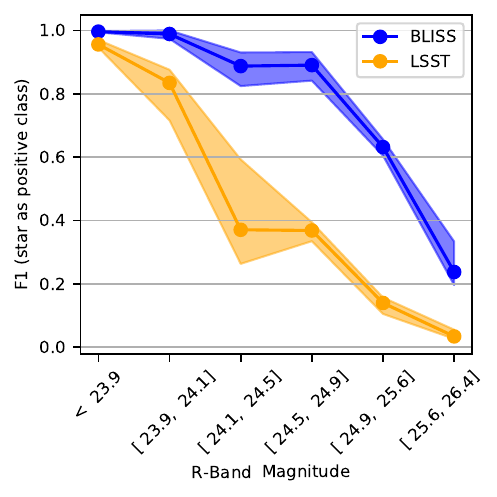}
    \caption{The F1 scores for star/galaxy classification, stratified by true object magnitude. On the left, galaxies serve as positive observations in defining the F1 score, whereas on the right, stars serve as positive observations.
    }
    \label{fig:classification_f1}
\end{figure}

As shown in \cref{fig:classification_f1}, BLISS consistently achieves better F1 scores than the LSST pipeline. With galaxies as the positive class, both BLISS and LSST attain F1 scores above 0.90 in all but the faintest magnitude bin. Conversely, with stars as the positive class, BLISS achieves F1 scores exceeding 0.90 in the four brightest bins, whereas LSST only exceeds this threshold in the brightest bin. 

Moreover, the F1 score deteriorates significantly with increasing magnitude. Specifically, the F1 score with stars as the positive class declines by 0.27 for BLISS from the second bin (magnitudes between 23.9 and 24.1) to the second-to-last bin (magnitudes between 24.9 and 25.6), while for LSST, it declines by 0.64. Thus, while both models perform better for brighter galaxies, the relative performance of BLISS increases greatly at fainter magnitudes.

\begin{figure}[ht]
    \centering
    \includegraphics[width=0.47\textwidth]{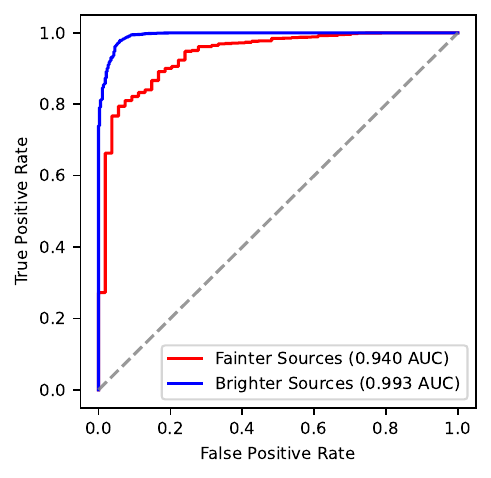}
    \caption{The ROC curve for star/galaxy classification by BLISS of brighter and fainter objects, split at 24.9 magnitude.  ``AUC'' refers to the area under the curve.}
    \label{fig:classification_roc}
\end{figure}

\paragraph*{ROC curve.}\quad
\cref{fig:classification_roc} further illustrates the difference in BLISS's classification performance for the brightest and the faintest objects by separating objects above and below magnitude 24.9. The area under the curve (AUC) for the brighter objects is 0.993 while the AUC for the fainter objects is 0.940. Brighter objects tend to have more distinct and well-defined features that help to distinguish between stars and galaxies.

\end{document}